\documentclass[journal=jacsat,manuscript=article]{achemso}
\usepackage{graphicx}
\usepackage{float}
\usepackage{amsmath}
\usepackage{subfigure}
\usepackage{braket}
\usepackage{color}
\usepackage[T1]{fontenc} 
\usepackage{afterpage}
\usepackage[dvipsnames]{xcolor}
\usepackage{epstopdf}
\usepackage[normalem]{ulem}
\usepackage{ulem}
\usepackage[T1]{fontenc}
\definecolor{ao(english)}{rgb}{0.0, 0.42, 0.24}

\newcommand{\cc}[1]{$\mathrm{g/cm^3}$}
\newcommand{\don}[1]{$\mathrm{n_{don}}$}
\newcommand{\acc}[1]{$\mathrm{n_{acc}}$}
\newcommand{\MC}[1]{\textcolor{black}{{#1}}}

\newcommand{\RX}[1]{\textcolor{black}{{#1}}}

\author{Renxi Liu}
\affiliation{HEDPS, CAPT, College of Engineering, Peking University, Beijing, 100871, P. R. China}
\alsoaffiliation{Academy for Advanced Interdisciplinary Studies, Peking University, Beijing, 90871, P. R. China}
\author{Mohan Chen}
\email{mohanchen@pku.edu.cn}
\affiliation{HEDPS, CAPT, College of Engineering, Peking University, Beijing, 100871, P. R. China}
\alsoaffiliation{Academy for Advanced Interdisciplinary Studies, Peking University, Beijing, 90871, P. R. China}
\altaffiliation{AI for Science Institute, Beijing 100080, P. R. China}
\date{\today}

\title{Characterization of the Hydrogen-Bond Network in High-Pressure Water by Deep Potential Molecular Dynamics}

\begin{document}
\begin{abstract}
%
%
%
%
%
%
%
The hydrogen-bond (H-bond) network of high-pressure water is investigated by neural-network-based molecular dynamics (MD) simulations with the first-principles accuracy. The static structure factors (SSFs) of water at three densities, i.e., 1, 1.115 and 1.24 \cc{} are directly evaluated from 512-water MD trajectories, which are in quantitative agreement with the experiments.
We propose a new method to decompose the computed SSF and identify the changes in SSF with respect to the changes in H-bond structures.
We find a larger water density results in a higher probability for one or two non-H-bonded water molecules to be inserted into the inner shell, explaining the changes in the tetrahedrality of water under pressure.
We predict that the structure of the accepting end of water molecules is more easily influenced by the pressure than the donating end. Our work sheds new light on explaining the SSF and H-bond properties in related fields. 
\end{abstract}

\section{Introduction}
Liquid water under high pressure has been the subject of intensive studies due to its importance in physics, chemistry, and life sciences.~\cite{06CSR-Daniel, 09B-French, 11JCP-Kang, 00L-Galli}
Among them, one particular focus is the distortion of the hydrogen bond (HB) network with respect to external pressure.
In the last two decades, 
a number of X-ray~\cite{94JCP-Okhulkov, 96MP-Radnai, 10B-Katayama, 12JML-Yamaguchi, 02JPCM-Eggert, 09B-Weck, 13PNAS-Sahle, 16JCP-Skinner} and neutron diffraction~\cite{00CP-Soper, 00L-Soper, 06L-Strassle} experiments have been conducted to determine the static structure factor (SSF) and radial distribution function (RDF)
of water under high pressures.
%
%
Through many efforts,
the influence of pressure on the SSF of liquid water is generally converged as follows.
Below 1 GPa, as pressure increases, the first peak of SSF rises significantly. In contrast, the second peak decreases monotonically in height, resulting in the fact that the highest second peak of SSF at ambient pressure shrinks into the shoulder of the first peak at 1 GPa.~\cite{16JCP-Skinner, 10B-Katayama, 02JPCM-Eggert}
The trend continues to higher pressures,
resulting in the fusion of the first two peaks into a single first peak at 4 GPa.~\cite{09B-Weck, 10B-Katayama, 12JML-Yamaguchi}
Recent works attribute the changes in the SSF to the damage of the tetrahedral structure of the inner shell~\cite{12JML-Yamaguchi, 16JCP-Skinner}.

Meanwhile, the RDF of oxygen atoms $\mathrm{g_{OO}(r)}$ also undergoes substantial changes as the pressure increases.
While the position of the first peak, which stands for the hydrogen-bonded water molecules, remains almost unchanged as pressure rises up to 1 GPa,~\cite{10B-Katayama, 00L-Soper, 00CP-Soper} 
the right shoulder of the first peak rises monotonically, and the second peak decreases to a local minimum.
Despite the converged experimental results,~\cite{16JCP-Skinner, 13JCP-Skinner} the influence of the hydrogen-bond network on the high-pressure water is still inconclusive, for instance, the following questions arise: (i) How do the changes in the HB network affect the first two peaks of SSF in the high-pressure water?
%
(ii) In terms of the number and the directionality of HBs, how does the pressure influence the accepting and donating ends of water molecules? 
(iii) How are the changes in the tetrahedral structure of water related to the HB network?
%


\afterpage{\begin{figure*}
  \includegraphics[width=15cm]{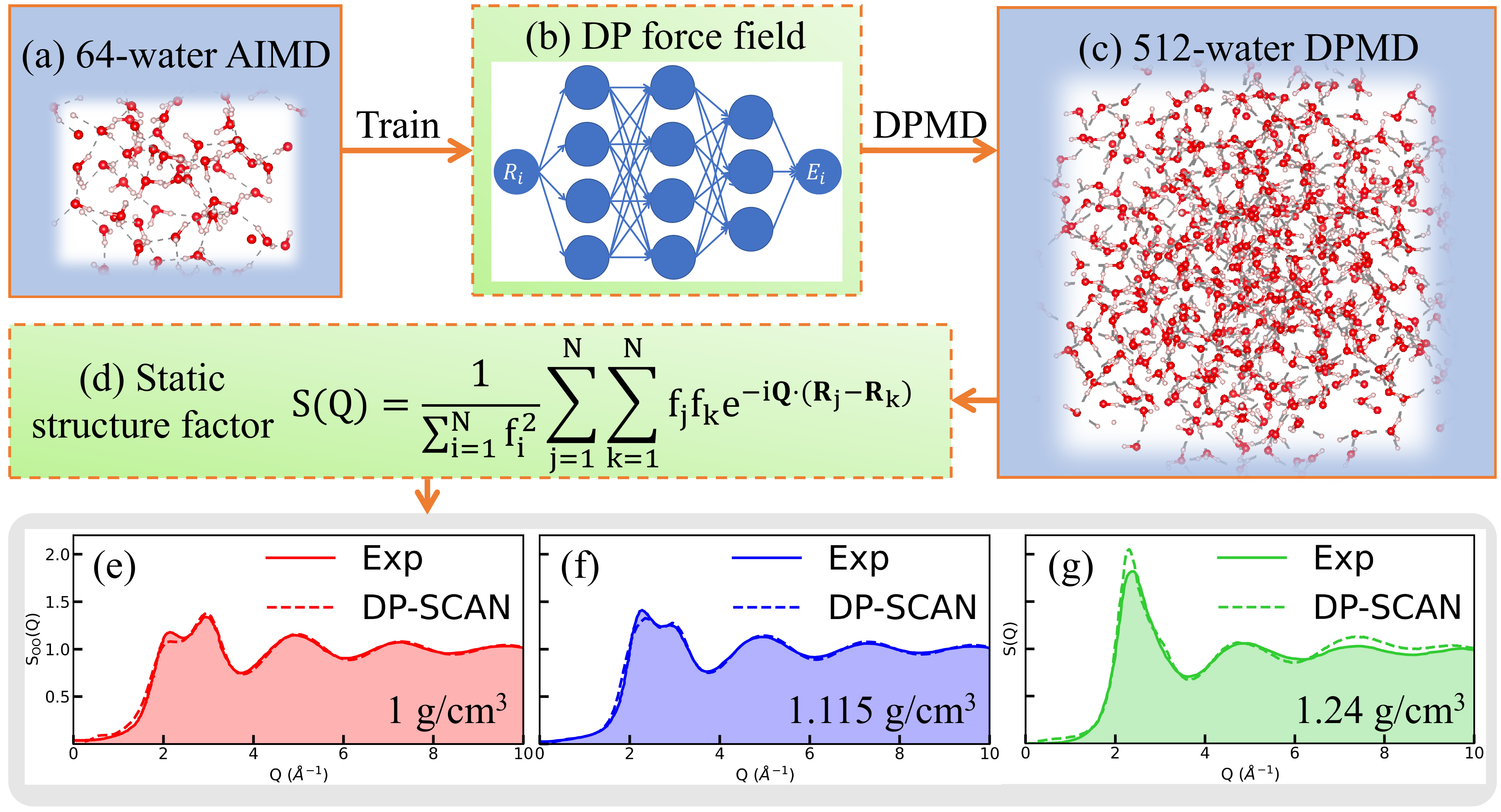}
  \caption{\MC{Workflow for calculations of static structure factors of liquid water at three densities (1, 1.115, 1.24 \cc{}). Training data are obtained from AIMD simulations of 64 water molecules.}
  (a-c) show the workflow of our calculation of static structure factor (SSF) under the three different pressures. 
  (d) shows the formula used to calculate the SSF, in which $f_i$ stands for the shape factor of atom $i$ in X-ray diffraction. For $\mathrm{S_{OO}(Q)}$ calculation, all $f_i$ are set to 1.
  (e-g) show the calculated SSFs compared with experiment results~\cite{13JCP-Skinner, 16JCP-Skinner, 12JML-Yamaguchi}, which in 1 and 1.115 \cc{} are $\mathrm{S_{OO}(Q)}$ and in 1.24 \cc{} is the total $\mathrm{S(Q)}$, since only total SSF in 1.24 \cc{} is provided in Ref.~\citenum{12JML-Yamaguchi}.} \label{workflow}
 \end{figure*}}

From a theoretical perspective,  
molecular dynamics (MD) serves as an important tool to investigate liquid water, bridging the gap between theory and experimentally measured SSF/RDF.
%
Past MD works~\cite{97JCP-Bagchi, 99E-Starr, 00L-Galli, 03E-Saitta, 06L-Strassle, 07E-Yan, 10JCP-Ikeda, 14JPCL-Fanetti} generally yielded that the typical tetrahedral HB structure is barely influenced by pressures under 1 GPa,
while the HB structure outside the first shell is changed and
occasionally one or two non-hydrogen-bonded water are pushed into the first shell.
In recent years, 
{\it ab initio} molecular dynamics (AIMD)~\cite{85L-CPMD} with first-principles accuracy is vastly applied in place of classical MD,
lifting the accuracy of simulation to a quantitative agreement with experiments.~\cite{16JCP-Skinner, 15PCCP-Imoto, 19JCP-Imoto, 19JPCB-Vondracek}
%
%
%
Although AIMD exhibits prediction power to explain the pressure effects on liquid water to a large extent,~\cite{15PCCP-Imoto, 16JCP-Skinner}
the small system size and short trajectory adopted by AIMD could only yield SSF with insufficient accuracy.
Besides, within the framework of density functional theory (DFT),~\cite{65PR-Kohn, 64PR-Hohenberg} as the \RX{generalized-gradient-approximation-level (GGA-level) exchange-correlation (XC)} functionals systematically overestimate the strength of the H-bonds and underestimate the density of water,
adding dispersion correction based on empirical parameters~\cite{10JCP-Grimme} has become a common practice in the modeling of water with GGA-level functionals,
thereby limiting the first-principles nature of AIMD.~\cite{09JPCB-Schmidt, 17PNAS-Chen, 11JCP-Wang, 14JCP-Distasio, 12JCP-Ma, 16PNAS-Morawietz}
In this regard, using AIMD to directly compute the SSF of liquid water remains a challenging issue.
%
%
%
%
\section{Methods}

\MC{
Our work aims to directly compute the static structure factors of liquid water at high pressures with first-principles accuracy and elucidate the related H-bond structures in detail.
%
%
We perform AIMD simulations with the meta-GGA exchange-correlation functional, SCAN,~\cite{15L-Sun}
which is known to satisfy all 17 restrictions on semi-local exchange-correlation functional and excellent for water properties.~\cite{16NC-Sun, 17PNAS-Chen, 18JCP-Zheng, 19B-Xu, 20PNAS-Sharkas, 21B-Xu, 22JCP-Liu}
For instance, the SCAN functional substantially improves the strong H-bonds description provided by GGA functionals such as PBE.
Meanwhile, the added intermediate-ranged van der Waals interactions in SCAN pull the second-shell water molecules closer to the interstitial area,
causing a denser and more disordered interstitial area around water molecules closer to the experimental data.~\cite{17PNAS-Chen, 18JCP-Zheng}
Consequently, the SCAN functional accurately predicts the H-bonded structure of liquid water.~\cite{17PNAS-Chen, 19B-Xu}
In this work, we perform three AIMD simulations of bulk water (64 water molecules) at ambient temperature with fixed densities of 1.0, 1.115, and 1.24 \cc{}, which correspond to water densities under ambient pressure, 360 MPa, and 1 GPa, respectively.~\cite{02JPCRD-Wagner}
}

\MC{A converged SSF needs MD simulations with a large system and a long trajectory, which is typically limited due to the high computational costs of AIMD.
In this regard, we adopt the Deep Potential Molecular Dynamics (DPMD)~\cite{18CPC-Wang, 18L-Zhang, 18CCP-Han} to train neural-network-based potentials for water with different densities; the workflow is illustrated in Figs.~\ref{workflow}(a)-(c).
The resulting neural networks are several orders of magnitude faster than AIMD and scale linearly with the number of atoms.
In previous works, DPMD has been validated to learn AIMD accuracy from both SCAN~\cite{21B-Xu, 20PNAS-Gartner, 21L-Zhang} and PBE0-TS functionals.~\cite{18L-Zhang}
In this work, the training sets, including the atomic positions, forces, and energy, are obtained from AIMD simulations with the SCAN functional in a 64-water cell. 
The energy predicted by DPMD is in quantitative agreement with the AIMD result (1 meV/atom).~\cite{18L-Zhang}
%
%
As a result, DPMD simulations are performed for 512 water molecules for 300 ps.
\RX{More details of the simulations and training can be found in Supplemental Material Section I.}
}

\section{Results and Discussions}
\MC{
Fig.~\ref{workflow}(d) lists the formula to compute SSF (S(Q)), and the results for the three densities are shown in Figs.~\ref{workflow}(e-g).
We find the DP model quantitatively reproduces the experimental data for all three densities.~\cite{13JCP-Skinner, 16JCP-Skinner, 12JML-Yamaguchi}
}
As the pressure increases from ambient to 1 GPa,
the first peak of $\mathrm{S_{OO}(Q)}$ significantly rises while the second peak decreases into a shoulder.
Correspondingly, 
as the position and shape of the first peak of $\mathrm{g_{OO}(r)}$ generally remain the same,
the second peak under ambient pressure decreases substantially as pressure rises.
The first minimum in between rises to form a bump beside the first peak.
The third peak also moves inward significantly as pressure rises.
Furthermore, we find that the height change of the first two peaks of SSF has nothing to do with the hydrogen-bonded first shell 
but purely results from the inward movement of non-hydrogen-bonded interstitial water molecules.
\begin{figure*}
  \includegraphics[width=15cm]{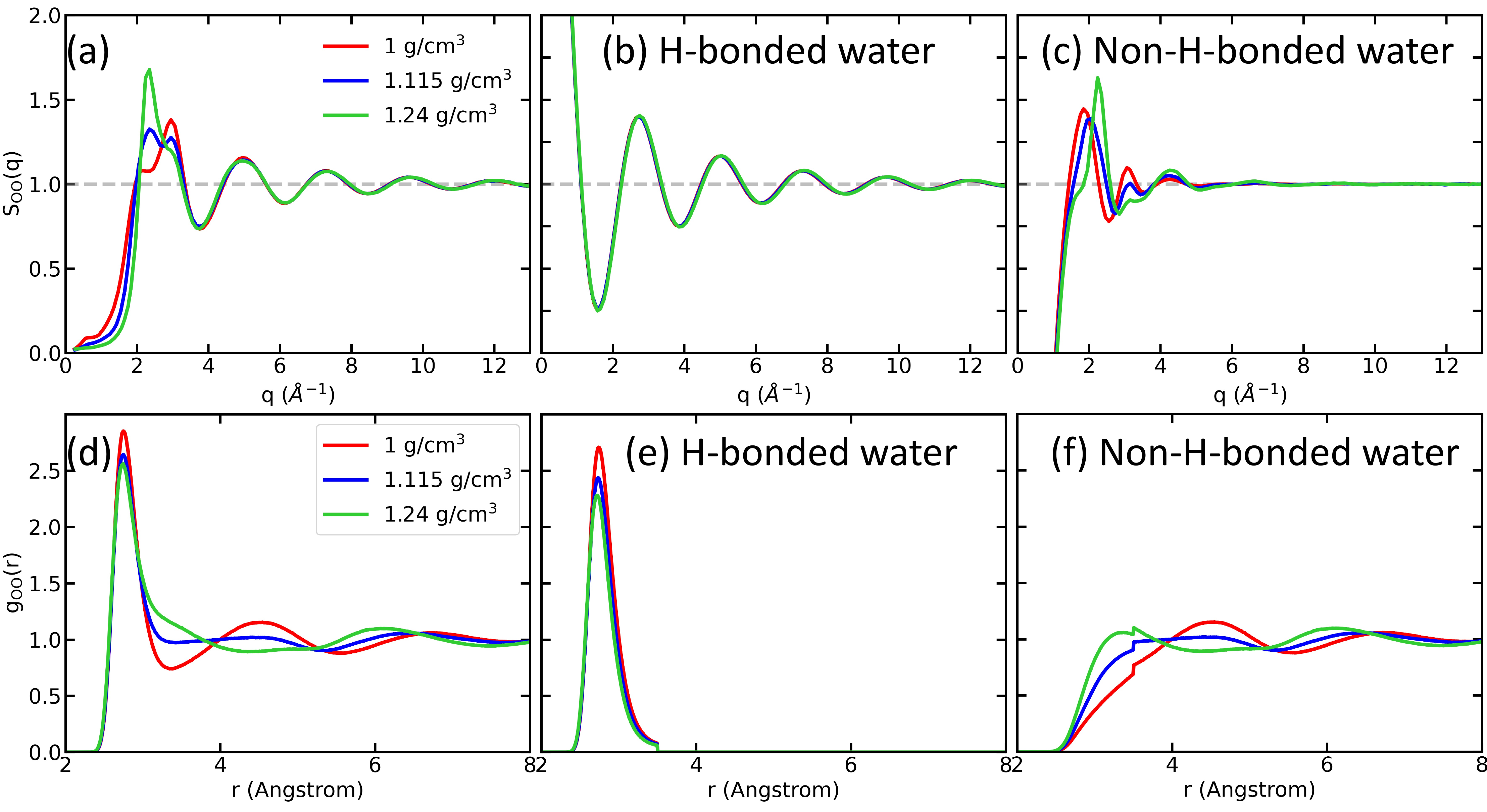}
  \caption{
  (a-c) Static structure factor (SSF) $\mathrm{S_{OO}(Q)}$, SSF contributed by hydrogen-bonded water molecule pairs, and SSF contributed by non-hydrogen-bonded water molecule pairs. (d-f) Radial distribution function (RDF) $\mathrm{g_{OO}(r)}$, RDF contributed by hydrogen-bonded water molecule pairs, and RDF contributed by non-hydrogen-bonded water molecule pairs. Water systems with a density of 1 \cc{}, 1.115 \cc{}, and 1.24 \cc{} are respectively colored in red, blue, and green.}\label{ssf}
 \end{figure*}
\MC{To further reveal the relations between the HBs in water and the detailed changes in SSF under pressure,
we decompose $\mathrm{S_{OO}(Q)}$ into contributions of different water molecule pairs with the formula as follows
\RX{
\begin{align}
\begin{split}
    S_{\mathrm{OO}}(\mathrm{Q}) &= \frac{1}{N}\sum_{j = 1}^{N}\sum_{k = 1}^{N}e^{i(\mathbf{r_j} - \mathbf{r_k})\cdot\mathbf{Q}} \\
    &= \frac{1}{N}\left\{N + \sum_{j = 1}^{N-1}\sum_{k = j+1}^{N}\Big[e^{i(\mathbf{r_j} - \mathbf{r_k})\cdot\mathbf{Q}} + e^{i(\mathbf{r_k} - \mathbf{r_j})\cdot\mathbf{Q}}\Big]\right\} \\
    &= \frac{1}{N}\left\{N + 2\sum_{j = 1}^{N-1}\sum_{k = j+1}^{N}\Big[\cos{(\mathbf{r_j} - \mathbf{r_k})\cdot\mathbf{Q}}\Big]\right\}\\
    &= 1+\frac{2}{N}\left\{\sum_{(j, k)\in HB}\cos{(\mathbf{r}_j-\mathbf{r}_k)\cdot \mathbf{Q}}+\sum_{(j, k)\notin HB}\cos{(\mathbf{r}_j-\mathbf{r}_k)\cdot \mathbf{Q}}\right\},
\end{split}
\end{align}
}
where $N$ stands for the number of atoms, $j$ and $k$ stand for the indices of atoms, and $\mathbf{r}$ stands for the position of atom.
Specifically, $\mathrm{S_{OO}(Q)}$ of water is decomposed into two components by classifying the atom pairs into hydrogen-bonded pair and non-hydrogen-bonded pairs, as displayed in Figs.~\ref{ssf}(b) and (c), respectively.
Notably, the HB criteria are chosen with the O-O distance less than 3.5~\AA~and the O-O-H angle smaller than $\mathrm{30^\circ}$;~\cite{96N-Luzar}
although new criteria for defining HBs have been proposed~\cite{15PCCP-Imoto, 10JPCB-Henchman, 14JCP-Gasparotto, 16JCTC-Gasparotto} for both ambient and high-pressure water systems, our conclusions are insensitive to the hydrogen-bond criteria, more details of which are shown in Supplemental Material Section IV.}
%
\MC{
We find the changes in $\mathrm{S_{OO}(Q)}$ across the three densities mainly come from the non-hydrogen-bonded water molecules.
On the one hand, Fig.~\ref{ssf}(b) shows that the three $\mathrm{S_{OO}(Q)}$ contributed by the hydrogen-bonded water molecules in the first solvation shell at different pressures are almost identical, 
indicating that $\mathrm{S_{OO}(Q)}$ is insensitive to the changes in the hydrogen-bonded water molecules when the density rises from 1 to 1.24 \cc{}.
On the other hand, the three $\mathrm{S_{OO}(Q)}$ contributed by the non-hydrogen-bonded water molecules, as illustrated in Fig.~\ref{ssf}(c), exhibit substantially different features especially when the wave vector $\mathrm{Q<4}$~\AA$^{-1}$.
In detail, the first peak in Fig.~\ref{ssf}(c) decreases slightly when the water density changes from 1.0 to 1.115 \cc{},
but increases from 1.115 to 1.24 \cc{}.
Meanwhile, the peak moves towards a larger wave vector $\mathrm{Q}$ when the density increases.
During such a rightward move, 
the peak gradually merges into the first peak of hydrogen-bonded $\mathrm{S_{OO}(Q)}$ at 2.7 \AA$^{-1}$, as illustrated in Fig.~\ref{ssf}(b).
In addition, we observe that the decrease of the second peak at 2.9 \AA$^{-1}$ in Fig.~\ref{ssf}(a) 
corresponds to the decrease of the small peak at the same wave vector in Fig.~\ref{ssf}(c), suggesting this experimental feature of $\mathrm{S_{OO}(Q)}$ is also related to the non-hydrogen-bonded water molecules.
In summary, the above changes in $\mathrm{S_{OO}(Q)}$ from both hydrogen-bonded and non-hydrogen-bonded water molecules fully explain the increase of the first peak and decrease of the second peak in $\mathrm{S_{OO}(Q)}$ (Fig.~\ref{ssf}(a)) when the density increases from 1.0 to 1.24 \cc{}.
}

%
\MC{The features of $\mathrm{S_{OO}(Q)}$ in Fig.~\ref{ssf}(a) can be further explained by its Fourier transform counterpart, which is the radial distribution function $\mathrm{g_{OO}(r)}$ shown in Fig.~\ref{ssf}(d).
By adopting the same criterion for decomposing $\mathrm{S_{OO}(Q)}$, we decompose $\mathrm{g_{OO}(r)}$ into hydrogen-bonded and non-hydrogen-bonded terms and the results are shown in Figs.~\ref{ssf}(e) and (f).
The main peaks in Fig.~\ref{ssf}(e) are contributed by hydrogen-bonded water molecules at different densities. We observe the unaffected positions of the three peaks as the density changes from 1.0 to 1.24 \cc{}, which is due to the relative preservation of local HB order.~\cite{10JCP-Ikeda}
We also notice that the amplitude of the peak decreases slightly at a larger density, which is caused solely by the increase in water density but not affected by the hydrogen-bonded tetrahedral structure.
%
Furthermore, as illustrated in Fig.~\ref{ssf}(f), we find the water molecules in the second and third shells move towards the inner shell under pressure, which is evidenced by the disappearance of the second peak at 4.6 \AA, and the leftward movement of the third peak from 6.7 to 6.1 \AA.
In particular, when the density increases from 1.0 to 1.24 \cc{}, the non-hydrogen-bonded water molecules in the second shell move towards the region of the first shell, filling the O-O void space, as evidenced by the emergence of the peak at 3.4 \AA. As a result, a shoulder of the first peak emerges in $\mathrm{g_{OO}(r)}$ when the density is 1.24 \cc{}, as shown in Fig.~\ref{ssf}(d).
In conclusion, the features of $\mathrm{S_{OO}(Q)}$ under pressure are mainly affected by the movements of the non-hydrogen-bonded water molecules from the interstitial region and beyond towards the inner hydrogen-bonded tetrahedral structure.
}
%

\begin{figure}
  \includegraphics[width=8.5cm]{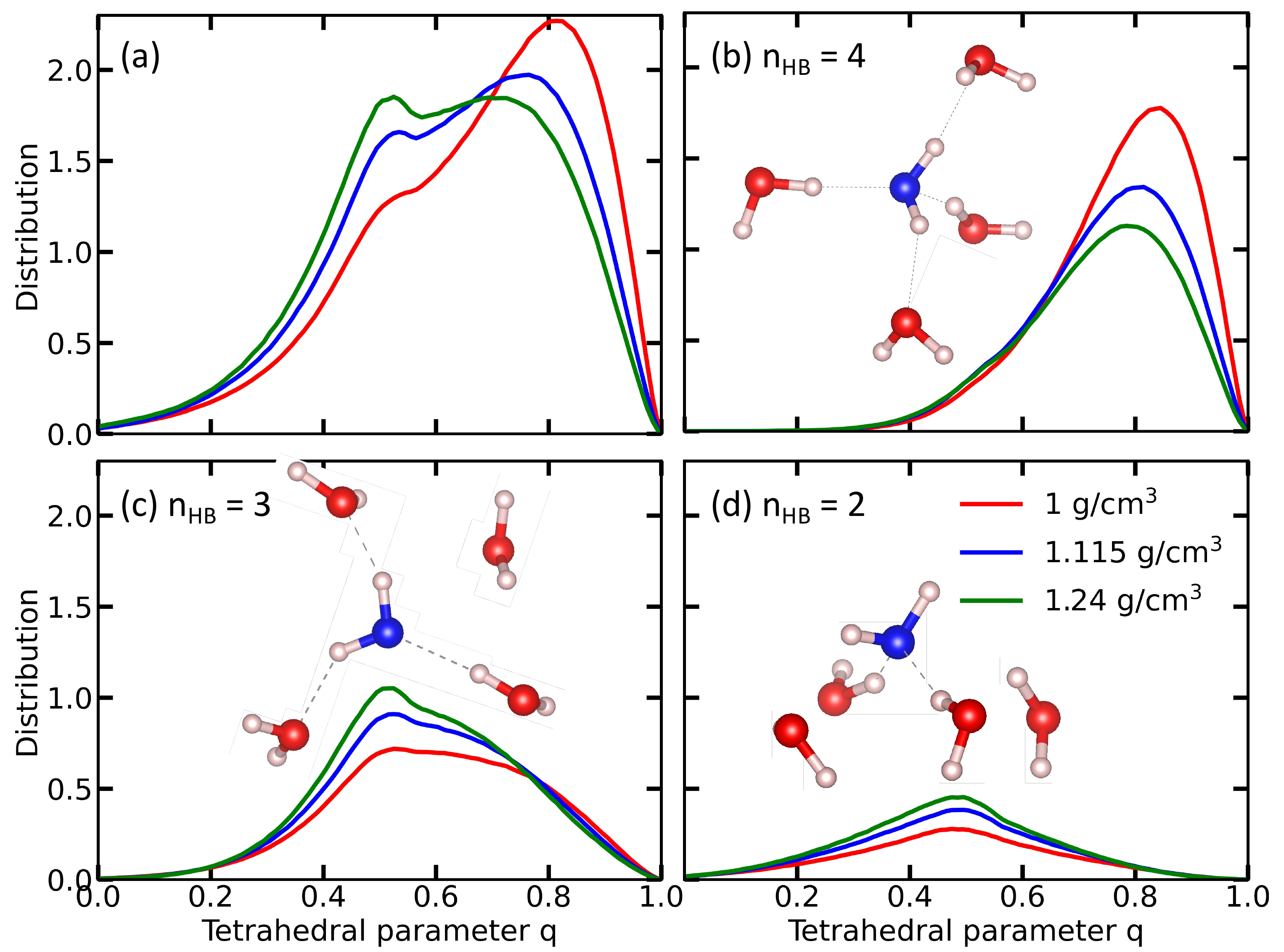}
  \caption{
  (a) Distribution of the tetrahedral parameter $q$ at 1 \cc{} (red), 1.115 \cc{} (blue) and 1.24 \cc{} (green). 
  (b-d) Distributions of $q$ contributed by water molecules whose 4, 3 and 2 neighbors out of the four closest neighbors are H-bonded, denoted by $\mathrm{n_{HB}=4, 3, 2}$, respectively.
  Inset in each subplot depicts a typical water with its four closest neighbors of the corresponding $\mathrm{n_{HB}}$.
  }\label{tetra}
 \end{figure}

\MC{
The $\mathrm{S_{OO}(Q)}$ and $\mathrm{g_{OO}(r)}$ yield HB information only in the radial direction. Since more experimental information is limited, first-principles simulations without empirical parameters play an important role in obtaining more HB network information for water under pressure.
Earlier literature reported that the H-bonded first shell of water molecules is almost unaffected, and the average number of HBs is not significantly influenced.~\cite{15PCCP-Imoto} However, when the water density changes from 1 to 1.24 \cc{}, the tetrahedral structure has been substantially altered according to our simulations.
Fig.~\ref{tetra}(a) illustrates the distributions of the tetrahedral parameter $q$ at the three densities from DPMD trajectories.~\cite{98MP-Chau, 01N-Errington}
The parameter $q$ measures the local structural order of liquid water in the sense of tetrahedrality.~\cite{98MP-Chau, 01N-Errington}
%
%
In our results,
$q$ distribution possesses a skewed peak at $q$=0.8 and a shoulder at $q$=0.5 under ambient pressure.
The height of the shoulder keeps rising while the peak decreases as the density changes from 1 to 1.24 \cc{},
forming a platform in the distribution between $q$=0.5 and $q$=0.8 at 1.24 \cc{},
which is in consistent with earlier studies.~\cite{10JCP-Ikeda, 16JCP-Skinner}
}
%
%
\MC{
To further investigate the changes in the distribution of the tetrahedral parameter $q$, we decompose the distribution by the number of H-bonded water molecules (denoted as $\mathrm{n_{HB}}$) within the four nearest neighbors of a water molecule.
The distributions contributed by $\mathrm{n_{HB}=}$4, 3, and 2 are shown in Figs.~\ref{tetra}(b), (c), and (d), respectively.
We find that the distribution involving four H-bonds exhibits a single peak at $q$=0.8 that represents a typical skewed tetrahedral HB structure in liquid water,
while the $\mathrm{n_{HB}=3}$ distribution exhibits a platform between $q$=0.5 and $q$=0.8, and $\mathrm{n_{HB}=2}$ forms a single-peak distribution centered at $q$=0.5.
As pressure increases, 
the percentage of $\mathrm{n_{HB}=4}$ drops substantially from 52.2\% (1 \cc{}) to 37.1\% (1.24 \cc{}) while percentages of $\mathrm{n_{HB}=2}$ and 3 increase,
resulting in the significant deformation of $q$.
}
%
%

\begin{figure}[htbp]
  \includegraphics[width=7.5cm]{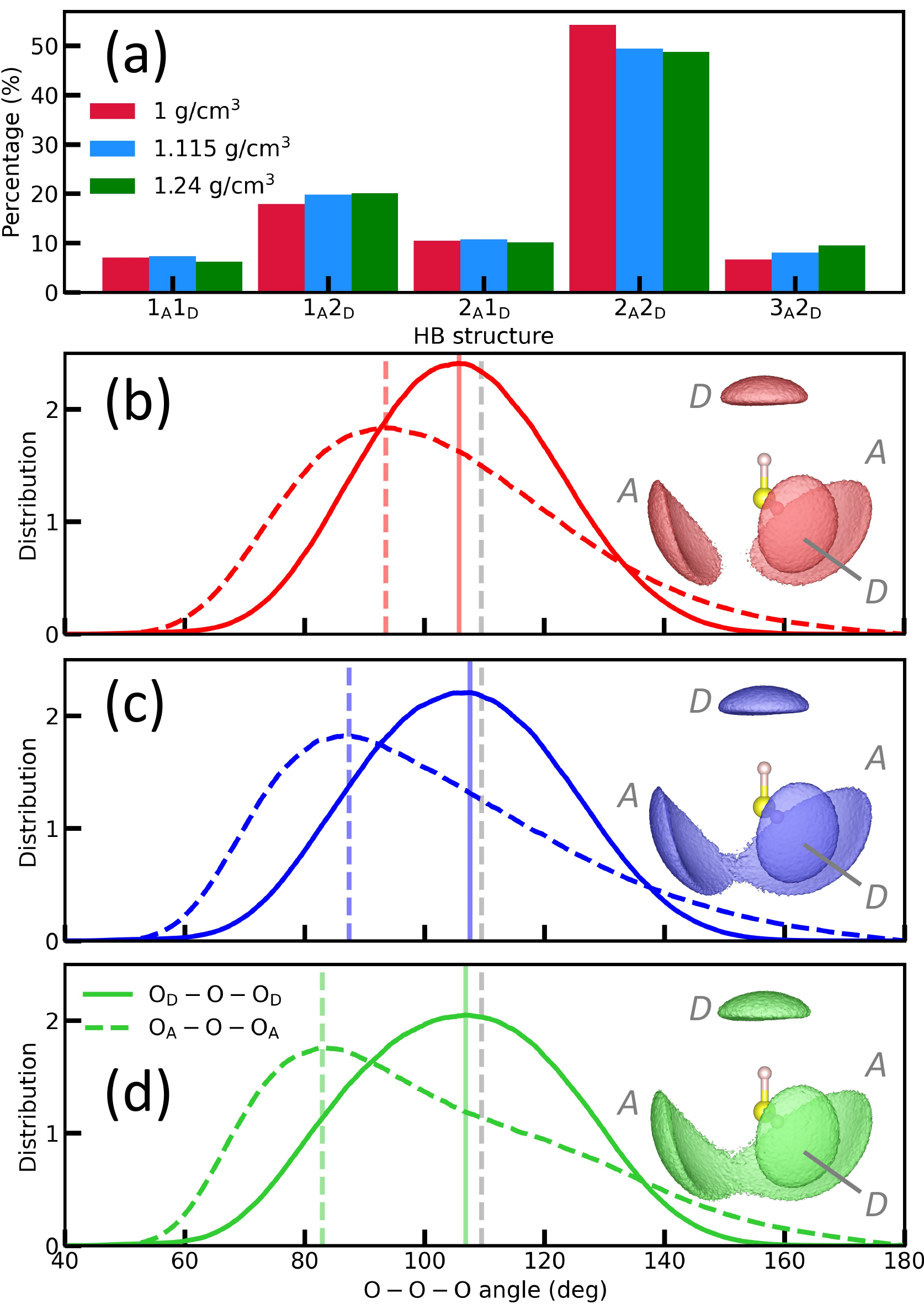}
  \caption{
  (a) Percentages of five major stable HB structures observed in liquid water with densities being 1, 1.115, and 1.24 \cc{}.
  (b-d) Distributions of the $\mathrm{O_A}$-$\mathrm{O}$-$\mathrm{O_A}$ angle (dashed line) and $\mathrm{O_D}$-$\mathrm{O}$-$\mathrm{O_D}$ angle (solid line) of the $\mathrm{2_A2_D}$ structure at the three different densities.
  Vertical dashed and solid lines in corresponding colors mark the peak positions of the $\mathrm{O_A}$-$\mathrm{O}$-$\mathrm{O_A}$ and $\mathrm{O_D}$-$\mathrm{O}$-$\mathrm{O_D}$ angle distribution, while the grey dashed line mark 109$^{\circ}$28$^{\prime}$.
  The three insets are the spatial distribution function (SDF) of the $\mathrm{2_A2_D}$ structure.
  }\label{hb_stru}
 \end{figure}

%
\MC{
Past studies attribute the decrease in the tetrahedrality to the insertion of one or two additional non-H-bonded water into the inner shell.~\cite{16JCP-Skinner}
However, our analysis shows that these structures already exist under ambient pressure.
Applying higher pressure does not create denser geometries at higher density but increases the probability of observing these structures.
Although a water molecule could accept or donate 1 to 3 hydrogen bonds in the liquid state,~\cite{11ACR-Agmon}
we only observe five main kinds of stable H-bond structures from the MD trajectories.
Fig.~\ref{hb_stru}(a) shows the classification of the five main H-bond structures, where a hydrogen-bond structure that accepts $m$ hydrogen bonds and donates $n$ hydrogen bonds is denoted as `$m_{\mathrm{A}}n_{\mathrm{D}}$'~\cite{16JCTC-Gasparotto}. 
We notice that the pressure mainly affects the accepting end rather than the donating end of water molecules.
For example, when the density rises from 1 to 1.24 \cc{}, the percentage of water molecules accepting 2 HBs substantially decreases from 64.7\% to 58.8\%,
while the percentage of donating 2 HBs only slightly decreases from 78.8\% to 78.3\%.
As a result, Fig.~\ref{hb_stru}(a) shows that the percentage of the $2_{\mathrm{A}}2_{\mathrm{D}}$ configurations decreases from 54.2\% at 1 \cc{} to 48.7\% at 1.24 \cc{}; around 5.5\% of water molecules deviate from the tetrahedral HB structure by gaining or losing an accepting HB. Consequently, the percentage of the $1_{\mathrm{A}}2_{\mathrm{D}}$ and $3_{\mathrm{A}}2_{\mathrm{D}}$ configurations increases monotonically as pressure increases.}
%

%

\MC{Pressure mainly affects the HB distribution of the accepting end of a water molecule rather than the donating end. Figs.~\ref{hb_stru}(b), (c), and (d) provide more details about the distributions of O-O-O angles for the tetrahdral $\mathrm{2_A2_D}$ water at a density of 1, 1.115, and 1.24 \cc{}, respectively. Here, the O-O-O angles are divided into two components as contributed by the accepting and donating ends. The statistics are based on the $\mathrm{2_A2_D}$ structure, which dominates HB structures at all three densities.
We further add in the O-O spatial distribution function (SDF), which provides a 3-dimensional view of the distribution of H-bonded O atoms close to the O of a water molecule.
At the donating end, we find the $\mathrm{O_D}$-$\mathrm{O}$-$\mathrm{O_D}$ angle distribution barely changes with pressure, as the peak position (vertical solid line in corresponding color) stays almost unchanged, close to the typical $\mathrm{109^{\circ}28^{'}}$ of tetrahedral structure (vertical grey line). As shown in the insets of Figs.~\ref{hb_stru}(b-d), the SDF forms two separate disk-like distributions and remains unchanged as pressure increases.
On the contrary, the $\mathrm{O_A}$-$\mathrm{O}$-$\mathrm{O_A}$ angle distribution moves leftward as pressure increases, the peak of which (dashed line in corresponding color) decreases from $\mathrm{93.6^{\circ}}$ to $\mathrm{83.0^{\circ}}$. Correspondingly, the inset of Fig.~\ref{hb_stru}(b) shows that the SDF at the accepting end exhibits two separate distributions at 1 \cc{}, but the two distributions spread more loosely and begin to merge at 1.115 \cc{}, and the trend becomes more clear at 1.24 \cc{}.
}

\RX{
We summarize our points as follows. 
A higher pressure decreases the percentage of standard tetrahedral HB structures and increases the percentage of non-tetrahedral HB structures, like $\mathrm{3_A2_D}$. 
For the tetrahedral $\mathrm{2_A2_D}$ HB structure, a higher pressure largely changes the $\mathrm{O_A-O-O_A}$ distribution on the accepting end but leaves the $\mathrm{O_D-O-O_D}$ distribution on the donating end almost unchanged (Fig. 4(b-d)). 
Increasing pressure also results in a higher chance of non-H-bonded water molecules moving into the inner shell, thereby further degrading the structural tetrahedrality. 
However, the average HB number per water molecule and the radial distribution of H-bonded water are hardly affected by the pressure. 
The increase in the density of the inner shell is mainly contributed by the increase in the non-H-bonded water inserted into the inner shell.
Overall, these features form a complete picture of the pressure influence on the structure of liquid water.  
}

\RX{
Furthermore, the deformation of the tetrahedral structure at the accepting end, along with the movement of water molecules from the interstitial region and beyond into the inner shell, could potentially result in a higher proportion of weak and distorted hydrogen bonds compared to water under ambient conditions.
The distortion of the HB geometry is also revealed by Fig.~S3 (c-e) in the Supplemental Material.
Since the nuclear quantum effects (NQEs) are suggested to generally strengthen the strong HBs, and weaken the weak ones,~\cite{11PNAS-Li, 16CR-Ceriotti}
it is likely that a higher number of weakened HBs and a more disordered HB network would be observed in water under high pressure when taking NQEs into account.
However, to what extent NQEs affect the structure of liquid water under pressure still requires further study involving the NQEs.
}

\section{Conclusions}
\MC{
In conclusion, we systematically investigated the changes in H-bond structures of liquid water under pressure by performing neural-network-based DPMD simulations with the first-principles accuracy. The use of machine-learning-assisted DPMD models largely boosted the efficiency of AIMD, enabling us to simulate a large cell with 512 water molecules and a trajectory length of 300 ps. Besides, the use of the SCAN functional provided first-principles accuracy for liquid water without empirical parameters. As a result, we 
directly computed the SSF of water at different densities and proposed a new method that links the H-bond structure information to the decomposed SSF. 
%
%
%
%
The results and analyses provided in this work may help better understand the influences of pressures on the H-bond network of water. Furthermore, it can be readily applied to study similar scientific problems that involve H-bond information in the X-ray or neutron diffraction experiments.
}

\begin{acknowledgement}

The work of R.L. and M.C. is supported by the National Science Foundation of China under Grant No. 12122401 and No. 12074007. All of the numerical simulations were performed on the High-Performance Computing Platform of CAPT and Bohrium platform supported by DP Technology. 

\end{acknowledgement}

\begin{suppinfo}
The following files are available as supplemental information free of charge.

\begin{itemize}
  \item SM.pdf (PDF): supplemental material including verification of the accuracy of the trained potential, robustness of the HB classification standard etc.
  \item SM.tex (LATEX): latex source file for compiling the SM.pdf file.
  \item SM\_ref.bib (LATEX): bibliography file for the SM.pdf file.
  \item FigureS1-S9.pdf (PDF): figures contained in SM.pdf file, details of which are elucidated in SM.pdf.
\end{itemize}

\end{suppinfo}

\bibliography{reference}

\providecommand{\latin}[1]{#1}
\makeatletter
\providecommand{\doi}
  {\begingroup\let\do\@makeother\dospecials
  \catcode`\{=1 \catcode`\}=2 \doi@aux}
\providecommand{\doi@aux}[1]{\endgroup\texttt{#1}}
\makeatother
\providecommand*\mcitethebibliography{\thebibliography}
\csname @ifundefined\endcsname{endmcitethebibliography}
  {\let\endmcitethebibliography\endthebibliography}{}
\begin{mcitethebibliography}{58}
\providecommand*\natexlab[1]{#1}
\providecommand*\mciteSetBstSublistMode[1]{}
\providecommand*\mciteSetBstMaxWidthForm[2]{}
\providecommand*\mciteBstWouldAddEndPuncttrue
  {\def\EndOfBibitem{\unskip.}}
\providecommand*\mciteBstWouldAddEndPunctfalse
  {\let\EndOfBibitem\relax}
\providecommand*\mciteSetBstMidEndSepPunct[3]{}
\providecommand*\mciteSetBstSublistLabelBeginEnd[3]{}
\providecommand*\EndOfBibitem{}
\mciteSetBstSublistMode{f}
\mciteSetBstMaxWidthForm{subitem}{(\alph{mcitesubitemcount})}
\mciteSetBstSublistLabelBeginEnd
  {\mcitemaxwidthsubitemform\space}
  {\relax}
  {\relax}

\bibitem[Daniel \latin{et~al.}(2006)Daniel, Oger, and Winter]{06CSR-Daniel}
Daniel,~I.; Oger,~P.; Winter,~R. Origins of life and biochemistry under
  high-pressure conditions. \emph{Chemical Society Reviews} \textbf{2006},
  \emph{35}, 858--875\relax
\mciteBstWouldAddEndPuncttrue
\mciteSetBstMidEndSepPunct{\mcitedefaultmidpunct}
{\mcitedefaultendpunct}{\mcitedefaultseppunct}\relax
\EndOfBibitem
\bibitem[French \latin{et~al.}(2009)French, Mattsson, Nettelmann, and
  Redmer]{09B-French}
French,~M.; Mattsson,~T.~R.; Nettelmann,~N.; Redmer,~R. Equation of state and
  phase diagram of water at ultrahigh pressures as in planetary interiors.
  \emph{Physical Review B} \textbf{2009}, \emph{79}, 054107\relax
\mciteBstWouldAddEndPuncttrue
\mciteSetBstMidEndSepPunct{\mcitedefaultmidpunct}
{\mcitedefaultendpunct}{\mcitedefaultseppunct}\relax
\EndOfBibitem
\bibitem[Kang \latin{et~al.}(2011)Kang, Dai, and Yuan]{11JCP-Kang}
Kang,~D.; Dai,~J.; Yuan,~J. Changes of structure and dipole moment of water
  with temperature and pressure: A first principles study. \emph{The Journal of
  Chemical Physics} \textbf{2011}, \emph{135}, 024505\relax
\mciteBstWouldAddEndPuncttrue
\mciteSetBstMidEndSepPunct{\mcitedefaultmidpunct}
{\mcitedefaultendpunct}{\mcitedefaultseppunct}\relax
\EndOfBibitem
\bibitem[Schwegler \latin{et~al.}(2000)Schwegler, Galli, and Gygi]{00L-Galli}
Schwegler,~E.; Galli,~G.; Gygi,~F.~G. Water under Pressure. \emph{Physical
  Review Letters} \textbf{2000}, \emph{84}, 2429--2432\relax
\mciteBstWouldAddEndPuncttrue
\mciteSetBstMidEndSepPunct{\mcitedefaultmidpunct}
{\mcitedefaultendpunct}{\mcitedefaultseppunct}\relax
\EndOfBibitem
\bibitem[Okhulkov \latin{et~al.}(1994)Okhulkov, Demianets, and
  Gorbaty]{94JCP-Okhulkov}
Okhulkov,~A.~V.; Demianets,~Y.~N.; Gorbaty,~Y.~E. X‐ray scattering in liquid
  water at pressures of up to 7.7 kbar: Test of a fluctuation model. \emph{The
  Journal of Chemical Physics} \textbf{1994}, \emph{100}, 1578--1588\relax
\mciteBstWouldAddEndPuncttrue
\mciteSetBstMidEndSepPunct{\mcitedefaultmidpunct}
{\mcitedefaultendpunct}{\mcitedefaultseppunct}\relax
\EndOfBibitem
\bibitem[Radnai and Ohtaki(1996)Radnai, and Ohtaki]{96MP-Radnai}
Radnai,~T.; Ohtaki,~H. X-ray diffraction studies on the structure of water at
  high temperatures and pressures. \emph{Molecular Physics} \textbf{1996},
  \emph{87}, 103--121\relax
\mciteBstWouldAddEndPuncttrue
\mciteSetBstMidEndSepPunct{\mcitedefaultmidpunct}
{\mcitedefaultendpunct}{\mcitedefaultseppunct}\relax
\EndOfBibitem
\bibitem[Katayama \latin{et~al.}(2010)Katayama, Hattori, Saitoh, Ikeda, Aoki,
  Fukui, and Funakoshi]{10B-Katayama}
Katayama,~Y.; Hattori,~T.; Saitoh,~H.; Ikeda,~T.; Aoki,~K.; Fukui,~H.;
  Funakoshi,~K. Structure of liquid water under high pressure up to 17 GPa.
  \emph{Physical Review B} \textbf{2010}, \emph{81}, 014109\relax
\mciteBstWouldAddEndPuncttrue
\mciteSetBstMidEndSepPunct{\mcitedefaultmidpunct}
{\mcitedefaultendpunct}{\mcitedefaultseppunct}\relax
\EndOfBibitem
\bibitem[Yamaguchi \latin{et~al.}(2012)Yamaguchi, Fujimura, Uchi, Yoshida, and
  Katayama]{12JML-Yamaguchi}
Yamaguchi,~T.; Fujimura,~K.; Uchi,~K.; Yoshida,~K.; Katayama,~Y. Structure of
  water from ambient to 4GPa revealed by energy-dispersive X-ray diffraction
  combined with empirical potential structure refinement modeling.
  \emph{Journal of Molecular Liquids} \textbf{2012}, \emph{176}, 44--51\relax
\mciteBstWouldAddEndPuncttrue
\mciteSetBstMidEndSepPunct{\mcitedefaultmidpunct}
{\mcitedefaultendpunct}{\mcitedefaultseppunct}\relax
\EndOfBibitem
\bibitem[Eggert \latin{et~al.}(2002)Eggert, Weck, and Loubeyre]{02JPCM-Eggert}
Eggert,~J.~H.; Weck,~G.; Loubeyre,~P. Structure of liquid water at high
  pressures and temperatures. \emph{Journal of Physics: Condensed Matter}
  \textbf{2002}, \emph{14}, 11385--11394\relax
\mciteBstWouldAddEndPuncttrue
\mciteSetBstMidEndSepPunct{\mcitedefaultmidpunct}
{\mcitedefaultendpunct}{\mcitedefaultseppunct}\relax
\EndOfBibitem
\bibitem[Weck \latin{et~al.}(2009)Weck, Eggert, Loubeyre, Desbiens, Bourasseau,
  Maillet, Mezouar, and Hanfland]{09B-Weck}
Weck,~G.; Eggert,~J.; Loubeyre,~P.; Desbiens,~N.; Bourasseau,~E.;
  Maillet,~J.-B.; Mezouar,~M.; Hanfland,~M. Phase diagrams and isotopic effects
  of normal and deuterated water studied via x-ray diffraction up to 4.5 GPa
  and 500 K. \emph{Physical Review B} \textbf{2009}, \emph{80}, 180202\relax
\mciteBstWouldAddEndPuncttrue
\mciteSetBstMidEndSepPunct{\mcitedefaultmidpunct}
{\mcitedefaultendpunct}{\mcitedefaultseppunct}\relax
\EndOfBibitem
\bibitem[Sahle \latin{et~al.}(2013)Sahle, Sternemann, Schmidt, Lehtola, Jahn,
  Simonelli, Huotari, Hakala, Pylkkänen, Nyrow, Mende, Tolan, Hämäläinen,
  and Wilke]{13PNAS-Sahle}
Sahle,~C.~J.; Sternemann,~C.; Schmidt,~C.; Lehtola,~S.; Jahn,~S.;
  Simonelli,~L.; Huotari,~S.; Hakala,~M.; Pylkkänen,~T.; Nyrow,~A.; Mende,~K.;
  Tolan,~M.; Hämäläinen,~K.; Wilke,~M. Microscopic structure of water at
  elevated pressures and temperatures. \emph{Proceedings of the National
  Academy of Sciences} \textbf{2013}, \emph{110}, 6301--6306\relax
\mciteBstWouldAddEndPuncttrue
\mciteSetBstMidEndSepPunct{\mcitedefaultmidpunct}
{\mcitedefaultendpunct}{\mcitedefaultseppunct}\relax
\EndOfBibitem
\bibitem[Skinner \latin{et~al.}(2016)Skinner, Galib, Fulton, Mundy, Parise,
  Pham, Schenter, and Benmore]{16JCP-Skinner}
Skinner,~L.~B.; Galib,~M.; Fulton,~J.~L.; Mundy,~C.~J.; Parise,~J.~B.;
  Pham,~V.-T.; Schenter,~G.~K.; Benmore,~C.~J. The structure of liquid water up
  to 360 MPa from x-ray diffraction measurements using a high Q-range and from
  molecular simulation. \emph{The Journal of Chemical Physics} \textbf{2016},
  \emph{144}, 134504\relax
\mciteBstWouldAddEndPuncttrue
\mciteSetBstMidEndSepPunct{\mcitedefaultmidpunct}
{\mcitedefaultendpunct}{\mcitedefaultseppunct}\relax
\EndOfBibitem
\bibitem[Soper(2000)]{00CP-Soper}
Soper,~A. The radial distribution functions of water and ice from 220 to 673 K
  and at pressures up to 400 MPa. \emph{Chemical Physics} \textbf{2000},
  \emph{258}, 121--137\relax
\mciteBstWouldAddEndPuncttrue
\mciteSetBstMidEndSepPunct{\mcitedefaultmidpunct}
{\mcitedefaultendpunct}{\mcitedefaultseppunct}\relax
\EndOfBibitem
\bibitem[Soper and Ricci(2000)Soper, and Ricci]{00L-Soper}
Soper,~A.~K.; Ricci,~M.~A. Structures of High-Density and Low-Density Water.
  \emph{Physical Review Letters} \textbf{2000}, \emph{84}, 2881--2884\relax
\mciteBstWouldAddEndPuncttrue
\mciteSetBstMidEndSepPunct{\mcitedefaultmidpunct}
{\mcitedefaultendpunct}{\mcitedefaultseppunct}\relax
\EndOfBibitem
\bibitem[Str\"assle \latin{et~al.}(2006)Str\"assle, Saitta, Godec, Hamel,
  Klotz, Loveday, and Nelmes]{06L-Strassle}
Str\"assle,~T.; Saitta,~A.~M.; Godec,~Y.~L.; Hamel,~G.; Klotz,~S.;
  Loveday,~J.~S.; Nelmes,~R.~J. Structure of Dense Liquid Water by Neutron
  Scattering to 6.5 GPa and 670 K. \emph{Physical Review Letters}
  \textbf{2006}, \emph{96}, 067801\relax
\mciteBstWouldAddEndPuncttrue
\mciteSetBstMidEndSepPunct{\mcitedefaultmidpunct}
{\mcitedefaultendpunct}{\mcitedefaultseppunct}\relax
\EndOfBibitem
\bibitem[Skinner \latin{et~al.}(2013)Skinner, Huang, Schlesinger, Pettersson,
  Nilsson, and Benmore]{13JCP-Skinner}
Skinner,~L.~B.; Huang,~C.; Schlesinger,~D.; Pettersson,~L. G.~M.; Nilsson,~A.;
  Benmore,~C.~J. Benchmark oxygen-oxygen pair-distribution function of ambient
  water from x-ray diffraction measurements with a wide Q-range. \emph{The
  Journal of Chemical Physics} \textbf{2013}, \emph{138}, 074506\relax
\mciteBstWouldAddEndPuncttrue
\mciteSetBstMidEndSepPunct{\mcitedefaultmidpunct}
{\mcitedefaultendpunct}{\mcitedefaultseppunct}\relax
\EndOfBibitem
\bibitem[Bagchi \latin{et~al.}(1997)Bagchi, Balasubramanian, and
  Klein]{97JCP-Bagchi}
Bagchi,~K.; Balasubramanian,~S.; Klein,~M.~L. The effects of pressure on
  structural and dynamical properties of associated liquids: Molecular dynamics
  calculations for the extended simple point charge model of water. \emph{The
  Journal of Chemical Physics} \textbf{1997}, \emph{107}, 8561--8567\relax
\mciteBstWouldAddEndPuncttrue
\mciteSetBstMidEndSepPunct{\mcitedefaultmidpunct}
{\mcitedefaultendpunct}{\mcitedefaultseppunct}\relax
\EndOfBibitem
\bibitem[Starr \latin{et~al.}(1999)Starr, Bellissent-Funel, and
  Stanley]{99E-Starr}
Starr,~F.~W.; Bellissent-Funel,~M.-C.; Stanley,~H.~E. Structure of supercooled
  and glassy water under pressure. \emph{Physical Review E} \textbf{1999},
  \emph{60}, 1084--1087\relax
\mciteBstWouldAddEndPuncttrue
\mciteSetBstMidEndSepPunct{\mcitedefaultmidpunct}
{\mcitedefaultendpunct}{\mcitedefaultseppunct}\relax
\EndOfBibitem
\bibitem[Saitta and Datchi(2003)Saitta, and Datchi]{03E-Saitta}
Saitta,~A.~M.; Datchi,~F. Structure and phase diagram of high-density water:
  The role of interstitial molecules. \emph{Physical Review E} \textbf{2003},
  \emph{67}, 020201\relax
\mciteBstWouldAddEndPuncttrue
\mciteSetBstMidEndSepPunct{\mcitedefaultmidpunct}
{\mcitedefaultendpunct}{\mcitedefaultseppunct}\relax
\EndOfBibitem
\bibitem[Yan \latin{et~al.}(2007)Yan, Buldyrev, Kumar, Giovambattista,
  Debenedetti, and Stanley]{07E-Yan}
Yan,~Z.; Buldyrev,~S.~V.; Kumar,~P.; Giovambattista,~N.; Debenedetti,~P.~G.;
  Stanley,~H.~E. Structure of the first- and second-neighbor shells of
  simulated water: Quantitative relation to translational and orientational
  order. \emph{Physical Review E} \textbf{2007}, \emph{76}, 051201\relax
\mciteBstWouldAddEndPuncttrue
\mciteSetBstMidEndSepPunct{\mcitedefaultmidpunct}
{\mcitedefaultendpunct}{\mcitedefaultseppunct}\relax
\EndOfBibitem
\bibitem[Ikeda \latin{et~al.}(2010)Ikeda, Katayama, Saitoh, and
  Aoki]{10JCP-Ikeda}
Ikeda,~T.; Katayama,~Y.; Saitoh,~H.; Aoki,~K. Communications: High-temperature
  water under pressure. \emph{The Journal of Chemical Physics} \textbf{2010},
  \emph{132}, 121102\relax
\mciteBstWouldAddEndPuncttrue
\mciteSetBstMidEndSepPunct{\mcitedefaultmidpunct}
{\mcitedefaultendpunct}{\mcitedefaultseppunct}\relax
\EndOfBibitem
\bibitem[Fanetti \latin{et~al.}(2014)Fanetti, Lapini, Pagliai, Citroni,
  Di~Donato, Scandolo, Righini, and Bini]{14JPCL-Fanetti}
Fanetti,~S.; Lapini,~A.; Pagliai,~M.; Citroni,~M.; Di~Donato,~M.; Scandolo,~S.;
  Righini,~R.; Bini,~R. Structure and Dynamics of Low-Density and High-Density
  Liquid Water at High Pressure. \emph{The Journal of Physical Chemistry
  Letters} \textbf{2014}, \emph{5}, 235--240\relax
\mciteBstWouldAddEndPuncttrue
\mciteSetBstMidEndSepPunct{\mcitedefaultmidpunct}
{\mcitedefaultendpunct}{\mcitedefaultseppunct}\relax
\EndOfBibitem
\bibitem[Car and Parrinello(1985)Car, and Parrinello]{85L-CPMD}
Car,~R.; Parrinello,~M. Unified Approach for Molecular Dynamics and
  Density-Functional Theory. \emph{Physical Review Letters} \textbf{1985},
  \emph{55}, 2471--2474\relax
\mciteBstWouldAddEndPuncttrue
\mciteSetBstMidEndSepPunct{\mcitedefaultmidpunct}
{\mcitedefaultendpunct}{\mcitedefaultseppunct}\relax
\EndOfBibitem
\bibitem[Imoto \latin{et~al.}(2015)Imoto, Forbert, and Marx]{15PCCP-Imoto}
Imoto,~S.; Forbert,~H.; Marx,~D. Water structure and solvation of osmolytes at
  high hydrostatic pressure: pure water and TMAO solutions at 10 kbar versus 1
  bar. \emph{Physical Chemistry Chemical Physics} \textbf{2015}, \emph{17},
  24224--24237\relax
\mciteBstWouldAddEndPuncttrue
\mciteSetBstMidEndSepPunct{\mcitedefaultmidpunct}
{\mcitedefaultendpunct}{\mcitedefaultseppunct}\relax
\EndOfBibitem
\bibitem[Imoto and Marx(2019)Imoto, and Marx]{19JCP-Imoto}
Imoto,~S.; Marx,~D. Pressure response of the THz spectrum of bulk liquid water
  revealed by intermolecular instantaneous normal mode analysis. \emph{The
  Journal of Chemical Physics} \textbf{2019}, \emph{150}, 084502\relax
\mciteBstWouldAddEndPuncttrue
\mciteSetBstMidEndSepPunct{\mcitedefaultmidpunct}
{\mcitedefaultendpunct}{\mcitedefaultseppunct}\relax
\EndOfBibitem
\bibitem[Vondracek \latin{et~al.}(2019)Vondracek, Imoto, Knake, Schwaab, Marx,
  and Havenith]{19JPCB-Vondracek}
Vondracek,~H.; Imoto,~S.; Knake,~L.; Schwaab,~G.; Marx,~D.; Havenith,~M.
  Hydrogen-Bonding in Liquid Water at Multikilobar Pressures. \emph{The Journal
  of Physical Chemistry B} \textbf{2019}, \emph{123}, 7748--7753\relax
\mciteBstWouldAddEndPuncttrue
\mciteSetBstMidEndSepPunct{\mcitedefaultmidpunct}
{\mcitedefaultendpunct}{\mcitedefaultseppunct}\relax
\EndOfBibitem
\bibitem[Kohn and Sham(1965)Kohn, and Sham]{65PR-Kohn}
Kohn,~W.; Sham,~L.~J. Self-Consistent Equations Including Exchange and
  Correlation Effects. \emph{Physical Review} \textbf{1965}, \emph{140},
  1133A\relax
\mciteBstWouldAddEndPuncttrue
\mciteSetBstMidEndSepPunct{\mcitedefaultmidpunct}
{\mcitedefaultendpunct}{\mcitedefaultseppunct}\relax
\EndOfBibitem
\bibitem[Hohenberg and Kohn(1964)Hohenberg, and Kohn]{64PR-Hohenberg}
Hohenberg,~P.; Kohn,~W. Inhomogeneous Electron Gas. \emph{Physical Review}
  \textbf{1964}, \emph{136}, 864B\relax
\mciteBstWouldAddEndPuncttrue
\mciteSetBstMidEndSepPunct{\mcitedefaultmidpunct}
{\mcitedefaultendpunct}{\mcitedefaultseppunct}\relax
\EndOfBibitem
\bibitem[Grimme \latin{et~al.}(2010)Grimme, Antony, Ehrlich, and
  Krieg]{10JCP-Grimme}
Grimme,~S.; Antony,~J.; Ehrlich,~S.; Krieg,~H. A consistent and accurate ab
  initio parametrization of density functional dispersion correction (DFT-D)
  for the 94 elements H-Pu. \emph{The Journal of Chemical Physics}
  \textbf{2010}, \emph{132}, 154104\relax
\mciteBstWouldAddEndPuncttrue
\mciteSetBstMidEndSepPunct{\mcitedefaultmidpunct}
{\mcitedefaultendpunct}{\mcitedefaultseppunct}\relax
\EndOfBibitem
\bibitem[Schmidt \latin{et~al.}(2009)Schmidt, VandeVondele, Kuo, Sebastiani,
  Siepmann, Hutter, and Mundy]{09JPCB-Schmidt}
Schmidt,~J.; VandeVondele,~J.; Kuo,~I. F.~W.; Sebastiani,~D.; Siepmann,~J.~I.;
  Hutter,~J.; Mundy,~C.~J. Isobaric-Isothermal Molecular Dynamics Simulations
  Utilizing Density Functional Theory: An Assessment of the Structure and
  Density of Water at Near-Ambient Conditions. \emph{The Journal of Physical
  Chemistry B} \textbf{2009}, \emph{113}, 11959--11964\relax
\mciteBstWouldAddEndPuncttrue
\mciteSetBstMidEndSepPunct{\mcitedefaultmidpunct}
{\mcitedefaultendpunct}{\mcitedefaultseppunct}\relax
\EndOfBibitem
\bibitem[Chen \latin{et~al.}(2017)Chen, Ko, Remsing, Calegari~Andrade, Santra,
  Sun, Selloni, Car, Klein, Perdew, and Wu]{17PNAS-Chen}
Chen,~M.; Ko,~H.-Y.; Remsing,~R.~C.; Calegari~Andrade,~M.~F.; Santra,~B.;
  Sun,~Z.; Selloni,~A.; Car,~R.; Klein,~M.~L.; Perdew,~J.~P.; Wu,~X. Ab initio
  theory and modeling of water. \emph{Proceedings of the National Academy of
  Sciences} \textbf{2017}, \emph{114}, 10846--10851\relax
\mciteBstWouldAddEndPuncttrue
\mciteSetBstMidEndSepPunct{\mcitedefaultmidpunct}
{\mcitedefaultendpunct}{\mcitedefaultseppunct}\relax
\EndOfBibitem
\bibitem[Wang \latin{et~al.}(2011)Wang, Rom{\'a}n-P{\'e}rez, Soler, Artacho,
  and Fern{\'a}ndez-Serra]{11JCP-Wang}
Wang,~J.; Rom{\'a}n-P{\'e}rez,~G.; Soler,~J.~M.; Artacho,~E.;
  Fern{\'a}ndez-Serra,~M.~V. Density, structure, and dynamics of water: The
  effect of van der Waals interactions. \emph{The Journal of Chemical Physics}
  \textbf{2011}, \emph{134}, 024516\relax
\mciteBstWouldAddEndPuncttrue
\mciteSetBstMidEndSepPunct{\mcitedefaultmidpunct}
{\mcitedefaultendpunct}{\mcitedefaultseppunct}\relax
\EndOfBibitem
\bibitem[DiStasio \latin{et~al.}(2014)DiStasio, Santra, Li, Wu, and
  Car]{14JCP-Distasio}
DiStasio,~R.~A.; Santra,~B.; Li,~Z.; Wu,~X.; Car,~R. The individual and
  collective effects of exact exchange and dispersion interactions on the ab
  initio structure of liquid water. \emph{The Journal of Chemical Physics}
  \textbf{2014}, \emph{141}, 084502\relax
\mciteBstWouldAddEndPuncttrue
\mciteSetBstMidEndSepPunct{\mcitedefaultmidpunct}
{\mcitedefaultendpunct}{\mcitedefaultseppunct}\relax
\EndOfBibitem
\bibitem[Ma \latin{et~al.}(2012)Ma, Zhang, and Tuckerman]{12JCP-Ma}
Ma,~Z.; Zhang,~Y.; Tuckerman,~M.~E. Ab initio molecular dynamics study of water
  at constant pressure using converged basis sets and empirical dispersion
  corrections. \emph{The Journal of Chemical Physics} \textbf{2012},
  \emph{137}, 044506\relax
\mciteBstWouldAddEndPuncttrue
\mciteSetBstMidEndSepPunct{\mcitedefaultmidpunct}
{\mcitedefaultendpunct}{\mcitedefaultseppunct}\relax
\EndOfBibitem
\bibitem[Morawietz \latin{et~al.}(2016)Morawietz, Singraber, Dellago, and
  Behler]{16PNAS-Morawietz}
Morawietz,~T.; Singraber,~A.; Dellago,~C.; Behler,~J. How van der Waals
  interactions determine the unique properties of water. \emph{Proceedings of
  the National Academy of Sciences} \textbf{2016}, \emph{113}, 8368--8373\relax
\mciteBstWouldAddEndPuncttrue
\mciteSetBstMidEndSepPunct{\mcitedefaultmidpunct}
{\mcitedefaultendpunct}{\mcitedefaultseppunct}\relax
\EndOfBibitem
\bibitem[Sun~J(2015)]{15L-Sun}
Sun~J,~P.~J.,~Ruzsinszky~A Strongly Constrained and Appropriately Normed
  Semilocal Density Functional. \emph{Physical Review Letters} \textbf{2015},
  \emph{115}, 036402\relax
\mciteBstWouldAddEndPuncttrue
\mciteSetBstMidEndSepPunct{\mcitedefaultmidpunct}
{\mcitedefaultendpunct}{\mcitedefaultseppunct}\relax
\EndOfBibitem
\bibitem[Sun \latin{et~al.}(2016)Sun, Remsing, Zhang, Sun, Ruzsinszky, Peng,
  Yang, Paul, Waghmare, Wu, Klein, and Perdew]{16NC-Sun}
Sun,~J.; Remsing,~R.~C.; Zhang,~Y.; Sun,~Z.; Ruzsinszky,~A.; Peng,~H.;
  Yang,~Z.; Paul,~A.; Waghmare,~U.; Wu,~X.; Klein,~M.~L.; Perdew,~J.~P.
  Accurate first-principles structures and energies of diversely bonded systems
  from an efficient density functional. \emph{Nature Chemistry} \textbf{2016},
  \emph{8}, 831--836\relax
\mciteBstWouldAddEndPuncttrue
\mciteSetBstMidEndSepPunct{\mcitedefaultmidpunct}
{\mcitedefaultendpunct}{\mcitedefaultseppunct}\relax
\EndOfBibitem
\bibitem[Zheng \latin{et~al.}(2018)Zheng, Chen, Sun, Ko, Santra, Dhuvad, and
  Wu]{18JCP-Zheng}
Zheng,~L.; Chen,~M.; Sun,~Z.; Ko,~H.-Y.; Santra,~B.; Dhuvad,~P.; Wu,~X.
  Structural, electronic, and dynamical properties of liquid water by ab initio
  molecular dynamics based on SCAN functional within the canonical ensemble.
  \emph{The Journal of Chemical Physics} \textbf{2018}, \emph{148},
  164505\relax
\mciteBstWouldAddEndPuncttrue
\mciteSetBstMidEndSepPunct{\mcitedefaultmidpunct}
{\mcitedefaultendpunct}{\mcitedefaultseppunct}\relax
\EndOfBibitem
\bibitem[Xu \latin{et~al.}(2019)Xu, Chen, Zhang, and Wu]{19B-Xu}
Xu,~J.; Chen,~M.; Zhang,~C.; Wu,~X. First-principles study of the infrared
  spectrum in liquid water from a systematically improved description of H-bond
  network. \emph{Physical Review B} \textbf{2019}, \emph{99}, 205123\relax
\mciteBstWouldAddEndPuncttrue
\mciteSetBstMidEndSepPunct{\mcitedefaultmidpunct}
{\mcitedefaultendpunct}{\mcitedefaultseppunct}\relax
\EndOfBibitem
\bibitem[Sharkas \latin{et~al.}(2020)Sharkas, Wagle, Santra, Akter, Zope,
  Baruah, Jackson, Perdew, and Peralta]{20PNAS-Sharkas}
Sharkas,~K.; Wagle,~K.; Santra,~B.; Akter,~S.; Zope,~R.~R.; Baruah,~T.;
  Jackson,~K.~A.; Perdew,~J.~P.; Peralta,~J.~E. Self-interaction error
  overbinds water clusters but cancels in structural energy differences.
  \emph{Proceedings of the National Academy of Sciences} \textbf{2020},
  \emph{117}, 11283--11288\relax
\mciteBstWouldAddEndPuncttrue
\mciteSetBstMidEndSepPunct{\mcitedefaultmidpunct}
{\mcitedefaultendpunct}{\mcitedefaultseppunct}\relax
\EndOfBibitem
\bibitem[Xu \latin{et~al.}(2020)Xu, Zhang, Zhang, Chen, Santra, and Wu]{21B-Xu}
Xu,~J.; Zhang,~C.; Zhang,~L.; Chen,~M.; Santra,~B.; Wu,~X. Isotope effects in
  molecular structures and electronic properties of liquid water via deep
  potential molecular dynamics based on the SCAN functional. \emph{Physical
  Review B} \textbf{2020}, \emph{102}, 214113\relax
\mciteBstWouldAddEndPuncttrue
\mciteSetBstMidEndSepPunct{\mcitedefaultmidpunct}
{\mcitedefaultendpunct}{\mcitedefaultseppunct}\relax
\EndOfBibitem
\bibitem[Liu \latin{et~al.}(2022)Liu, Zhang, Liang, Liu, Wu, and
  Chen]{22JCP-Liu}
Liu,~R.; Zhang,~C.; Liang,~X.; Liu,~J.; Wu,~X.; Chen,~M. Structural and dynamic
  properties of solvated hydroxide and hydronium ions in water from ab initio
  modeling. \emph{The Journal of Chemical Physics} \textbf{2022}, \emph{157},
  024503\relax
\mciteBstWouldAddEndPuncttrue
\mciteSetBstMidEndSepPunct{\mcitedefaultmidpunct}
{\mcitedefaultendpunct}{\mcitedefaultseppunct}\relax
\EndOfBibitem
\bibitem[Wagner and Pru{\ss}(2002)Wagner, and Pru{\ss}]{02JPCRD-Wagner}
Wagner,~W.; Pru{\ss},~A. The IAPWS Formulation 1995 for the Thermodynamic
  Properties of Ordinary Water Substance for General and Scientific Use.
  \emph{Journal of Physical and Chemical Reference Data} \textbf{2002},
  \emph{31}, 387--535\relax
\mciteBstWouldAddEndPuncttrue
\mciteSetBstMidEndSepPunct{\mcitedefaultmidpunct}
{\mcitedefaultendpunct}{\mcitedefaultseppunct}\relax
\EndOfBibitem
\bibitem[Wang \latin{et~al.}(2018)Wang, Zhang, Han, and E]{18CPC-Wang}
Wang,~H.; Zhang,~L.; Han,~J.; E,~W. DeePMD-kit: A deep learning package for
  many-body potential energy representation and molecular dynamics.
  \emph{Computer Physics Communications} \textbf{2018}, \emph{228},
  178--184\relax
\mciteBstWouldAddEndPuncttrue
\mciteSetBstMidEndSepPunct{\mcitedefaultmidpunct}
{\mcitedefaultendpunct}{\mcitedefaultseppunct}\relax
\EndOfBibitem
\bibitem[Zhang \latin{et~al.}(2018)Zhang, Han, Wang, Car, and E]{18L-Zhang}
Zhang,~L.; Han,~J.; Wang,~H.; Car,~R.; E,~W. Deep Potential Molecular Dynamics:
  A Scalable Model with the Accuracy of Quantum Mechanics. \emph{Physical
  Review Letters} \textbf{2018}, \emph{120}, 143001\relax
\mciteBstWouldAddEndPuncttrue
\mciteSetBstMidEndSepPunct{\mcitedefaultmidpunct}
{\mcitedefaultendpunct}{\mcitedefaultseppunct}\relax
\EndOfBibitem
\bibitem[Jiequn~Han(2018)]{18CCP-Han}
Jiequn~Han,~R. C. . W.~E.,~Linfeng~Zhang Deep Potential: A General
  Representation of a Many-Body Potential Energy Surface. \emph{Communications
  in Computational Physics} \textbf{2018}, \emph{23}, 629--639\relax
\mciteBstWouldAddEndPuncttrue
\mciteSetBstMidEndSepPunct{\mcitedefaultmidpunct}
{\mcitedefaultendpunct}{\mcitedefaultseppunct}\relax
\EndOfBibitem
\bibitem[Gartner \latin{et~al.}(2020)Gartner, Zhang, Piaggi, Car,
  Panagiotopoulos, and Debenedetti]{20PNAS-Gartner}
Gartner,~T.~E.; Zhang,~L.; Piaggi,~P.~M.; Car,~R.; Panagiotopoulos,~A.~Z.;
  Debenedetti,~P.~G. Signatures of a liquid{\textendash}liquid transition in an
  ab initio deep neural network model for water. \emph{Proceedings of the
  National Academy of Sciences} \textbf{2020}, \emph{117}, 26040--26046\relax
\mciteBstWouldAddEndPuncttrue
\mciteSetBstMidEndSepPunct{\mcitedefaultmidpunct}
{\mcitedefaultendpunct}{\mcitedefaultseppunct}\relax
\EndOfBibitem
\bibitem[Zhang \latin{et~al.}(2021)Zhang, Wang, Car, and E]{21L-Zhang}
Zhang,~L.; Wang,~H.; Car,~R.; E,~W. Phase Diagram of a Deep Potential Water
  Model. \emph{Physical Review Letters} \textbf{2021}, \emph{126}, 236001\relax
\mciteBstWouldAddEndPuncttrue
\mciteSetBstMidEndSepPunct{\mcitedefaultmidpunct}
{\mcitedefaultendpunct}{\mcitedefaultseppunct}\relax
\EndOfBibitem
\bibitem[Luzar and Chandler(1996)Luzar, and Chandler]{96N-Luzar}
Luzar,~A.; Chandler,~D. Hydrogen-bond kinetics in liquid water. \emph{Nature}
  \textbf{1996}, \emph{379}, 55--57\relax
\mciteBstWouldAddEndPuncttrue
\mciteSetBstMidEndSepPunct{\mcitedefaultmidpunct}
{\mcitedefaultendpunct}{\mcitedefaultseppunct}\relax
\EndOfBibitem
\bibitem[Henchman and Irudayam(2010)Henchman, and Irudayam]{10JPCB-Henchman}
Henchman,~R.~H.; Irudayam,~S.~J. Topological Hydrogen-Bond Definition to
  Characterize the Structure and Dynamics of Liquid Water. \emph{The Journal of
  Physical Chemistry B} \textbf{2010}, \emph{114}, 16792--16810\relax
\mciteBstWouldAddEndPuncttrue
\mciteSetBstMidEndSepPunct{\mcitedefaultmidpunct}
{\mcitedefaultendpunct}{\mcitedefaultseppunct}\relax
\EndOfBibitem
\bibitem[Gasparotto and Ceriotti(2014)Gasparotto, and
  Ceriotti]{14JCP-Gasparotto}
Gasparotto,~P.; Ceriotti,~M. Recognizing molecular patterns by machine
  learning: An agnostic structural definition of the hydrogen bond. \emph{The
  Journal of Chemical Physics} \textbf{2014}, \emph{141}, 174110\relax
\mciteBstWouldAddEndPuncttrue
\mciteSetBstMidEndSepPunct{\mcitedefaultmidpunct}
{\mcitedefaultendpunct}{\mcitedefaultseppunct}\relax
\EndOfBibitem
\bibitem[Gasparotto \latin{et~al.}(2016)Gasparotto, Hassanali, and
  Ceriotti]{16JCTC-Gasparotto}
Gasparotto,~P.; Hassanali,~A.~A.; Ceriotti,~M. Probing Defects and Correlations
  in the Hydrogen-Bond Network of ab Initio Water. \emph{Journal of Chemical
  Theory and Computation} \textbf{2016}, \emph{12}, 1953--1964\relax
\mciteBstWouldAddEndPuncttrue
\mciteSetBstMidEndSepPunct{\mcitedefaultmidpunct}
{\mcitedefaultendpunct}{\mcitedefaultseppunct}\relax
\EndOfBibitem
\bibitem[CHAU and HARDWICK(1998)CHAU, and HARDWICK]{98MP-Chau}
CHAU,~P.~L.; HARDWICK,~A.~J. A new order parameter for tetrahedral
  configurations. \emph{Molecular Physics} \textbf{1998}, \emph{93},
  511--518\relax
\mciteBstWouldAddEndPuncttrue
\mciteSetBstMidEndSepPunct{\mcitedefaultmidpunct}
{\mcitedefaultendpunct}{\mcitedefaultseppunct}\relax
\EndOfBibitem
\bibitem[Errington and Debenedetti(2001)Errington, and
  Debenedetti]{01N-Errington}
Errington,~J.~R.; Debenedetti,~P.~G. Relationship between structural order and
  the anomalies of liquid water. \emph{Nature} \textbf{2001}, \emph{409},
  318--321\relax
\mciteBstWouldAddEndPuncttrue
\mciteSetBstMidEndSepPunct{\mcitedefaultmidpunct}
{\mcitedefaultendpunct}{\mcitedefaultseppunct}\relax
\EndOfBibitem
\bibitem[Agmon(2012)]{11ACR-Agmon}
Agmon,~N. Liquid Water: From Symmetry Distortions to Diffusive Motion.
  \emph{Accounts of Chemical Research} \textbf{2012}, \emph{45}, 63--73\relax
\mciteBstWouldAddEndPuncttrue
\mciteSetBstMidEndSepPunct{\mcitedefaultmidpunct}
{\mcitedefaultendpunct}{\mcitedefaultseppunct}\relax
\EndOfBibitem
\bibitem[Li \latin{et~al.}(2011)Li, Walker, and Michaelides]{11PNAS-Li}
Li,~X.-Z.; Walker,~B.; Michaelides,~A. Quantum nature of the hydrogen bond.
  \emph{Proceedings of the National Academy of Sciences} \textbf{2011},
  \emph{108}, 6369--6373\relax
\mciteBstWouldAddEndPuncttrue
\mciteSetBstMidEndSepPunct{\mcitedefaultmidpunct}
{\mcitedefaultendpunct}{\mcitedefaultseppunct}\relax
\EndOfBibitem
\bibitem[Ceriotti \latin{et~al.}(2016)Ceriotti, Fang, Kusalik, McKenzie,
  Michaelides, Morales, and Markland]{16CR-Ceriotti}
Ceriotti,~M.; Fang,~W.; Kusalik,~P.~G.; McKenzie,~R.~H.; Michaelides,~A.;
  Morales,~M.~A.; Markland,~T.~E. Nuclear Quantum Effects in Water and Aqueous
  Systems: Experiment, Theory, and Current Challenges. \emph{Chemical Reviews}
  \textbf{2016}, \emph{116}, 7529--7550\relax
\mciteBstWouldAddEndPuncttrue
\mciteSetBstMidEndSepPunct{\mcitedefaultmidpunct}
{\mcitedefaultendpunct}{\mcitedefaultseppunct}\relax
\EndOfBibitem
\end{mcitethebibliography}


\providecommand{\latin}[1]{#1}
\makeatletter
\providecommand{\doi}
  {\begingroup\let\do\@makeother\dospecials
  \catcode`\{=1 \catcode`\}=2 \doi@aux}
\providecommand{\doi@aux}[1]{\endgroup\texttt{#1}}
\makeatother
\providecommand*\mcitethebibliography{\thebibliography}
\csname @ifundefined\endcsname{endmcitethebibliography}
  {\let\endmcitethebibliography\endthebibliography}{}
\begin{mcitethebibliography}{20}
\providecommand*\natexlab[1]{#1}
\providecommand*\mciteSetBstSublistMode[1]{}
\providecommand*\mciteSetBstMaxWidthForm[2]{}
\providecommand*\mciteBstWouldAddEndPuncttrue
  {\def\EndOfBibitem{\unskip.}}
\providecommand*\mciteBstWouldAddEndPunctfalse
  {\let\EndOfBibitem\relax}
\providecommand*\mciteSetBstMidEndSepPunct[3]{}
\providecommand*\mciteSetBstSublistLabelBeginEnd[3]{}
\providecommand*\EndOfBibitem{}
\mciteSetBstSublistMode{f}
\mciteSetBstMaxWidthForm{subitem}{(\alph{mcitesubitemcount})}
\mciteSetBstSublistLabelBeginEnd
  {\mcitemaxwidthsubitemform\space}
  {\relax}
  {\relax}

\bibitem[Car and Parrinello(1985)Car, and Parrinello]{85L-CPMD}
Car,~R.; Parrinello,~M. Unified Approach for Molecular Dynamics and
  Density-Functional Theory. \emph{Physical Review Letters} \textbf{1985},
  \emph{55}, 2471--2474\relax
\mciteBstWouldAddEndPuncttrue
\mciteSetBstMidEndSepPunct{\mcitedefaultmidpunct}
{\mcitedefaultendpunct}{\mcitedefaultseppunct}\relax
\EndOfBibitem
\bibitem[Giannozzi \latin{et~al.}(2009)Giannozzi, Baroni, Bonini, Calandra,
  Car, Cavazzoni, Ceresoli, Chiarotti, Cococcioni, Dabo, Corso, de~Gironcoli,
  Fabris, Fratesi, Gebauer, Gerstmann, Gougoussis, Kokalj, Lazzeri,
  Martin-Samos, Marzari, Mauri, Mazzarello, Paolini, Pasquarello, Paulatto,
  Sbraccia, Scandolo, Sclauzero, PSeitsonen, Smogunov, Umari, and
  Wentzcovitch]{codeQE}
Giannozzi,~P. \latin{et~al.}  QUANTUM ESPRESSO: a modular and open-source
  software project for quantum simulations of materials. \emph{Journal of
  Physics: Condensed Matter} \textbf{2009}, \emph{21}, 395502\relax
\mciteBstWouldAddEndPuncttrue
\mciteSetBstMidEndSepPunct{\mcitedefaultmidpunct}
{\mcitedefaultendpunct}{\mcitedefaultseppunct}\relax
\EndOfBibitem
\bibitem[Hamann \latin{et~al.}(1979)Hamann, Schl\"uter, and Chiang]{79L-Hamann}
Hamann,~D.~R.; Schl\"uter,~M.; Chiang,~C. Norm-Conserving Pseudopotentials.
  \emph{Physical Review Letters} \textbf{1979}, \emph{43}, 1494--1497\relax
\mciteBstWouldAddEndPuncttrue
\mciteSetBstMidEndSepPunct{\mcitedefaultmidpunct}
{\mcitedefaultendpunct}{\mcitedefaultseppunct}\relax
\EndOfBibitem
\bibitem[Nos\'{e}(1984)]{84JCP-Nose}
Nos\'{e},~S. A unified formulation of the constant temperature molecular
  dynamics methods. \emph{Journal of Chemical Physics} \textbf{1984},
  \emph{81}, 511--519\relax
\mciteBstWouldAddEndPuncttrue
\mciteSetBstMidEndSepPunct{\mcitedefaultmidpunct}
{\mcitedefaultendpunct}{\mcitedefaultseppunct}\relax
\EndOfBibitem
\bibitem[Hoover(1985)]{85A-Hoover}
Hoover,~W.~G. Canonical dynamics: Equilibrium phase-space distributions.
  \emph{Physical Review A} \textbf{1985}, \emph{31}, 1695--1697\relax
\mciteBstWouldAddEndPuncttrue
\mciteSetBstMidEndSepPunct{\mcitedefaultmidpunct}
{\mcitedefaultendpunct}{\mcitedefaultseppunct}\relax
\EndOfBibitem
\bibitem[Martyna \latin{et~al.}(1992)Martyna, Klein, and
  Tuckerman]{92JCP-Martyna}
Martyna,~G.~J.; Klein,~M.~L.; Tuckerman,~M. Nos\'{e}-Hoover chains: The
  canonical ensemble via continuous dynamics. \emph{Journal of Physical
  Chemistry} \textbf{1992}, \emph{97}, 2635--2643\relax
\mciteBstWouldAddEndPuncttrue
\mciteSetBstMidEndSepPunct{\mcitedefaultmidpunct}
{\mcitedefaultendpunct}{\mcitedefaultseppunct}\relax
\EndOfBibitem
\bibitem[Thompson \latin{et~al.}(2022)Thompson, Aktulga, Berger, Bolintineanu,
  Brown, Crozier, in~'t Veld, Kohlmeyer, Moore, Nguyen, Shan, Stevens,
  Tranchida, Trott, and Plimpton]{22CPC-lammps}
Thompson,~A.~P.; Aktulga,~H.~M.; Berger,~R.; Bolintineanu,~D.~S.; Brown,~W.~M.;
  Crozier,~P.~S.; in~'t Veld,~P.~J.; Kohlmeyer,~A.; Moore,~S.~G.;
  Nguyen,~T.~D.; Shan,~R.; Stevens,~M.~J.; Tranchida,~J.; Trott,~C.;
  Plimpton,~S.~J. LAMMPS - a flexible simulation tool for particle-based
  materials modeling at the atomic, meso, and continuum scales. \emph{Computer
  Physics Communications} \textbf{2022}, \emph{271}, 108171\relax
\mciteBstWouldAddEndPuncttrue
\mciteSetBstMidEndSepPunct{\mcitedefaultmidpunct}
{\mcitedefaultendpunct}{\mcitedefaultseppunct}\relax
\EndOfBibitem
\bibitem[Skinner \latin{et~al.}(2013)Skinner, Huang, Schlesinger, Pettersson,
  Nilsson, and Benmore]{13JCP-Skinner}
Skinner,~L.~B.; Huang,~C.; Schlesinger,~D.; Pettersson,~L. G.~M.; Nilsson,~A.;
  Benmore,~C.~J. Benchmark oxygen-oxygen pair-distribution function of ambient
  water from x-ray diffraction measurements with a wide Q-range. \emph{The
  Journal of Chemical Physics} \textbf{2013}, \emph{138}, 074506\relax
\mciteBstWouldAddEndPuncttrue
\mciteSetBstMidEndSepPunct{\mcitedefaultmidpunct}
{\mcitedefaultendpunct}{\mcitedefaultseppunct}\relax
\EndOfBibitem
\bibitem[Skinner \latin{et~al.}(2016)Skinner, Galib, Fulton, Mundy, Parise,
  Pham, Schenter, and Benmore]{16JCP-Skinner}
Skinner,~L.~B.; Galib,~M.; Fulton,~J.~L.; Mundy,~C.~J.; Parise,~J.~B.;
  Pham,~V.-T.; Schenter,~G.~K.; Benmore,~C.~J. The structure of liquid water up
  to 360 MPa from x-ray diffraction measurements using a high Q-range and from
  molecular simulation. \emph{The Journal of Chemical Physics} \textbf{2016},
  \emph{144}, 134504\relax
\mciteBstWouldAddEndPuncttrue
\mciteSetBstMidEndSepPunct{\mcitedefaultmidpunct}
{\mcitedefaultendpunct}{\mcitedefaultseppunct}\relax
\EndOfBibitem
\bibitem[Yamaguchi \latin{et~al.}(2012)Yamaguchi, Fujimura, Uchi, Yoshida, and
  Katayama]{12JML-Yamaguchi}
Yamaguchi,~T.; Fujimura,~K.; Uchi,~K.; Yoshida,~K.; Katayama,~Y. Structure of
  water from ambient to 4GPa revealed by energy-dispersive X-ray diffraction
  combined with empirical potential structure refinement modeling.
  \emph{Journal of Molecular Liquids} \textbf{2012}, \emph{176}, 44--51\relax
\mciteBstWouldAddEndPuncttrue
\mciteSetBstMidEndSepPunct{\mcitedefaultmidpunct}
{\mcitedefaultendpunct}{\mcitedefaultseppunct}\relax
\EndOfBibitem
\bibitem[Luzar and Chandler(1996)Luzar, and Chandler]{96N-Luzar}
Luzar,~A.; Chandler,~D. Hydrogen-bond kinetics in liquid water. \emph{Nature}
  \textbf{1996}, \emph{379}, 55--57\relax
\mciteBstWouldAddEndPuncttrue
\mciteSetBstMidEndSepPunct{\mcitedefaultmidpunct}
{\mcitedefaultendpunct}{\mcitedefaultseppunct}\relax
\EndOfBibitem
\bibitem[Laage and Hynes(2006)Laage, and Hynes]{06S-Laage}
Laage,~D.; Hynes,~J.~T. A Molecular Jump Mechanism of Water Reorientation.
  \emph{Science} \textbf{2006}, \emph{311}, 832--835\relax
\mciteBstWouldAddEndPuncttrue
\mciteSetBstMidEndSepPunct{\mcitedefaultmidpunct}
{\mcitedefaultendpunct}{\mcitedefaultseppunct}\relax
\EndOfBibitem
\bibitem[Errington and Debenedetti(2001)Errington, and
  Debenedetti]{01N-Errington}
Errington,~J.~R.; Debenedetti,~P.~G. Relationship between structural order and
  the anomalies of liquid water. \emph{Nature} \textbf{2001}, \emph{409},
  318--321\relax
\mciteBstWouldAddEndPuncttrue
\mciteSetBstMidEndSepPunct{\mcitedefaultmidpunct}
{\mcitedefaultendpunct}{\mcitedefaultseppunct}\relax
\EndOfBibitem
\bibitem[Berendsen \latin{et~al.}(1987)Berendsen, Grigera, and
  Straatsma]{87JPC-Berendson}
Berendsen,~H. J.~C.; Grigera,~J.~R.; Straatsma,~T.~P. The missing term in
  effective pair potentials. \emph{The Journal of Physical Chemistry}
  \textbf{1987}, \emph{91}, 6269--6271\relax
\mciteBstWouldAddEndPuncttrue
\mciteSetBstMidEndSepPunct{\mcitedefaultmidpunct}
{\mcitedefaultendpunct}{\mcitedefaultseppunct}\relax
\EndOfBibitem
\bibitem[Bagchi \latin{et~al.}(1997)Bagchi, Balasubramanian, and
  Klein]{97JCP-Bagchi}
Bagchi,~K.; Balasubramanian,~S.; Klein,~M.~L. {The effects of pressure on
  structural and dynamical properties of associated liquids: Molecular dynamics
  calculations for the extended simple point charge model of water}. \emph{The
  Journal of Chemical Physics} \textbf{1997}, \emph{107}, 8561--8567\relax
\mciteBstWouldAddEndPuncttrue
\mciteSetBstMidEndSepPunct{\mcitedefaultmidpunct}
{\mcitedefaultendpunct}{\mcitedefaultseppunct}\relax
\EndOfBibitem
\bibitem[Saitta and Datchi(2003)Saitta, and Datchi]{03E-Saitta}
Saitta,~A.~M.; Datchi,~F. Structure and phase diagram of high-density water:
  The role of interstitial molecules. \emph{Physical Review E} \textbf{2003},
  \emph{67}, 020201\relax
\mciteBstWouldAddEndPuncttrue
\mciteSetBstMidEndSepPunct{\mcitedefaultmidpunct}
{\mcitedefaultendpunct}{\mcitedefaultseppunct}\relax
\EndOfBibitem
\bibitem[Markovitch and Agmon(2008)Markovitch, and Agmon]{08MP-Markovitch}
Markovitch,~O.; Agmon,~N. The distribution of acceptor and donor hydrogen-bonds
  in bulk liquid water. \emph{Molecular Physics} \textbf{2008}, \emph{106},
  485--495\relax
\mciteBstWouldAddEndPuncttrue
\mciteSetBstMidEndSepPunct{\mcitedefaultmidpunct}
{\mcitedefaultendpunct}{\mcitedefaultseppunct}\relax
\EndOfBibitem
\bibitem[Marzari and Vanderbilt(1997)Marzari, and Vanderbilt]{97B-Marzari}
Marzari,~N.; Vanderbilt,~D. Maximally localized generalized Wannier functions
  for composite energy bands. \emph{Physical Review B} \textbf{1997},
  \emph{56}, 12847--12865\relax
\mciteBstWouldAddEndPuncttrue
\mciteSetBstMidEndSepPunct{\mcitedefaultmidpunct}
{\mcitedefaultendpunct}{\mcitedefaultseppunct}\relax
\EndOfBibitem
\bibitem[Marzari \latin{et~al.}(2012)Marzari, Mostofi, Yates, Souza, and
  Vanderbilt]{12RMP-Marzari}
Marzari,~N.; Mostofi,~A.~A.; Yates,~J.~R.; Souza,~I.; Vanderbilt,~D. Maximally
  localized Wannier functions: Theory and applications. \emph{Reviews of Modern
  Physics} \textbf{2012}, \emph{84}, 1419--1475\relax
\mciteBstWouldAddEndPuncttrue
\mciteSetBstMidEndSepPunct{\mcitedefaultmidpunct}
{\mcitedefaultendpunct}{\mcitedefaultseppunct}\relax
\EndOfBibitem
\end{mcitethebibliography}
\begin{figure*}[htbp]
  \includegraphics[width=9cm]{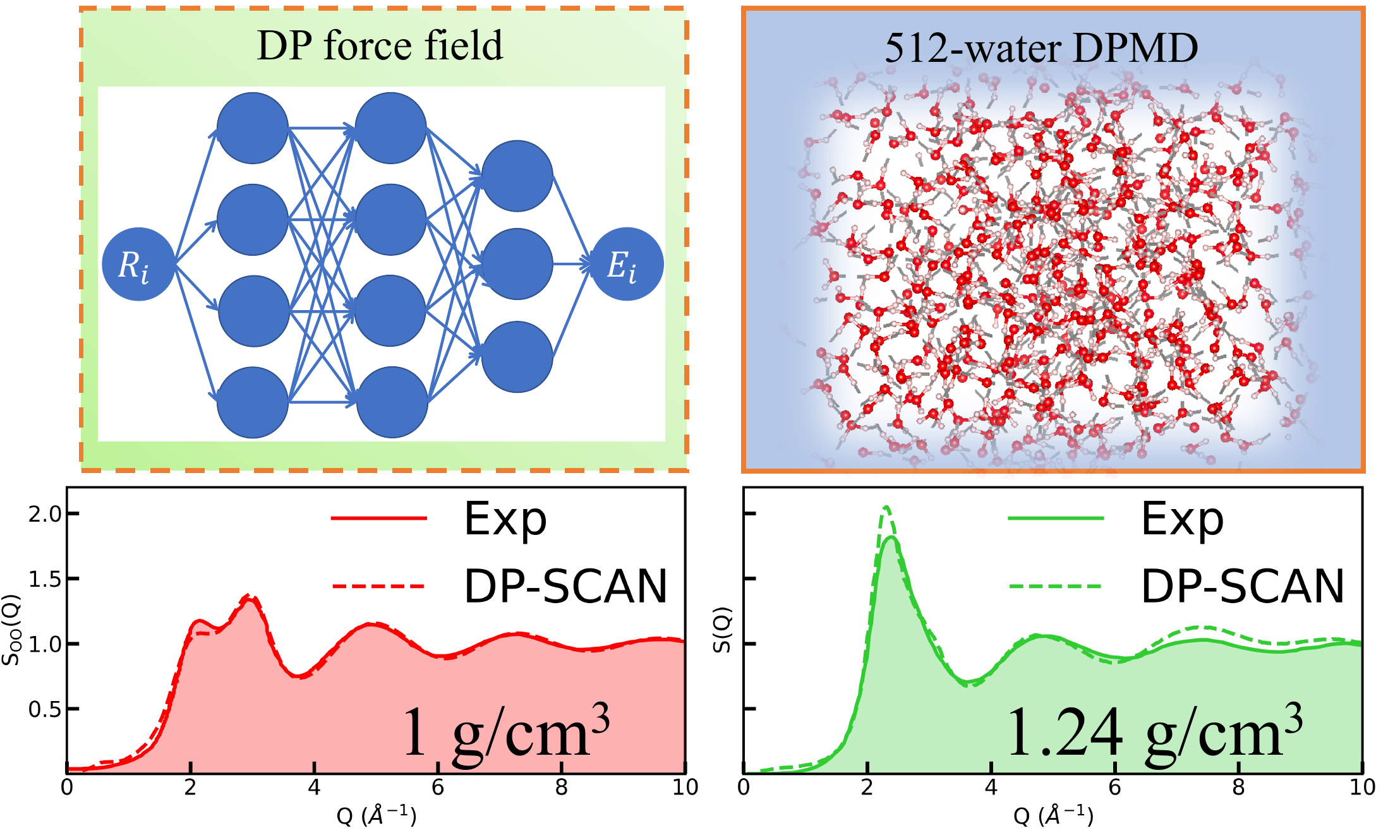}
  \caption*{
 For Table of Contents Only.
  }\label{toc}
 \end{figure*}

\end{document}


\section{Computational Details}

\RX{The AIMD simulations were performed using the Car-Parrinello molecular dynamics method~\cite{85L-CPMD} implemented in the Quantum ESPRESSO package.~\cite{codeQE}
A system consisting of 64 water molecules was simulated in periodic cubic boxes. The cell length was 12.4447 \AA, 11.9716 \AA~and 11.5550~\AA~for systems with a density of 1.0 \cc{}, 1.115 \cc{} and 1.24 \cc{}, respectively.
A kinetic energy cutoff of 85 Ry was applied, together with the SCAN meta-GGA XC functional and the Hamann-Schl$\ddot{u}$ter-Chiang-Vanderbilt pseudopotentials
generated with the PBE XC functional.~\cite{79L-Hamann}
We used the Nos\'{e}-Hoover chain thermostats with a chain length of 4 for each ion to control the ionic temperature of 330 K in the NVT ensemble.~\cite{84JCP-Nose, 85A-Hoover, 92JCP-Martyna}
We set the mass of hydrogen ion to be 2.0141 (mass of deuterium).
The fictitious mass of electron was set to be 100 a. u., and a mass cutoff of 25 a. u. was used.
The MD trajectory lengths were 30.8, 56.2, and 57.1 ps for the systems with densities being 1.0, 1.115, and 1.24 $\mathrm{g/cm^{3}}$, respectively. 
The MD time step was set to be 2 a. u., 
and MD information including atomic positions, forces were output every 5 MD steps.}

\RX{For the DP training, three DP models were trained separately, corresponding to the three different densities.
All output snapshots in the MD trajectories without the first 5 ps were used as the training data.
The smoothing radial cutoff was set to be 5.8~\AA~and the radial cutoff was set to be 6~\AA.
Embedding nets were all fully-connected neural networks with 3 hidden layers having 25, 50 and 100 neurons.
Fitting nets were fully-connected neural networks with 3 hidden layers having 240 neurons respectively.
The three models were all trained for 1,000,000 batches to ensure convergence.
The DPMD simulations were performed for 512 water molecules at the three densities for 300 ps with the LAMMPS software.~\cite{22CPC-lammps} 
The MD timestep was set to be 0.1 fs.
The Nos\'{e}-Hoover chain thermostats were used with a temperature damping parameter of 10 fs to control the temperature at 330 K.
}

\section{Accuracy of DPMD Models}

\begin{figure}[htp]
  \includegraphics[width=12cm]{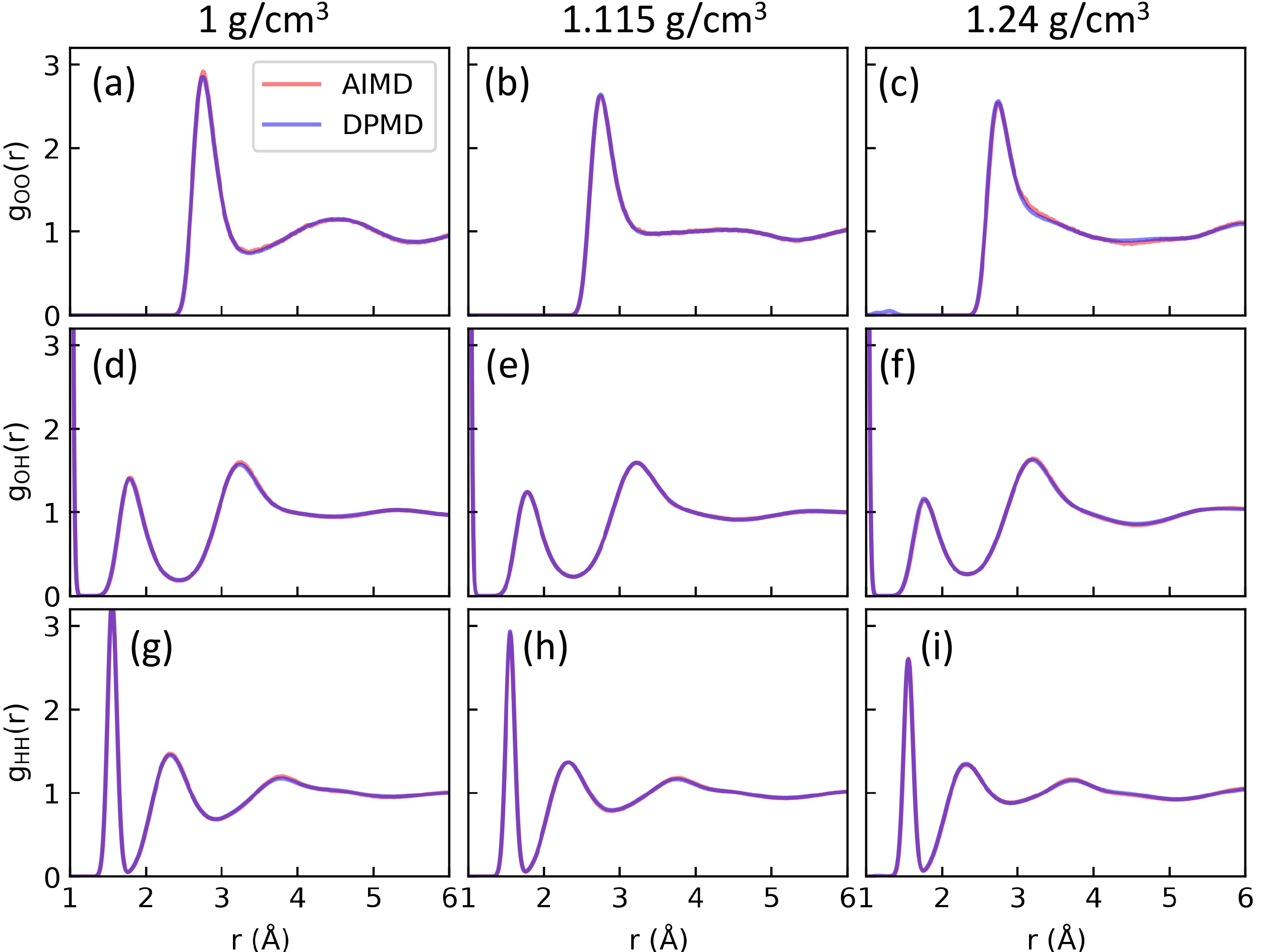}
  \caption{
  Comparison of RDFs $\mathrm{g_{OO}(r)}$, $\mathrm{g_{OH}(r)}$ and $\mathrm{g_{HH}(r)}$ between AIMD (red) and DPMD (blue) at 1, 1.115, and 1.24 \cc{}.
  The three rows from top to bottom correspond to $\mathrm{g_{OO}(r)}$, $\mathrm{g_{OH}(r)}$ and $\mathrm{g_{HH}(r)}$, respectively.
  The three columns from left to right correspond to 1, 1.115, and 1.24 \cc{}, respectively.
  }\label{pdf_compare}
 \end{figure}

\begin{figure}[htp]
  \includegraphics[width=14cm]{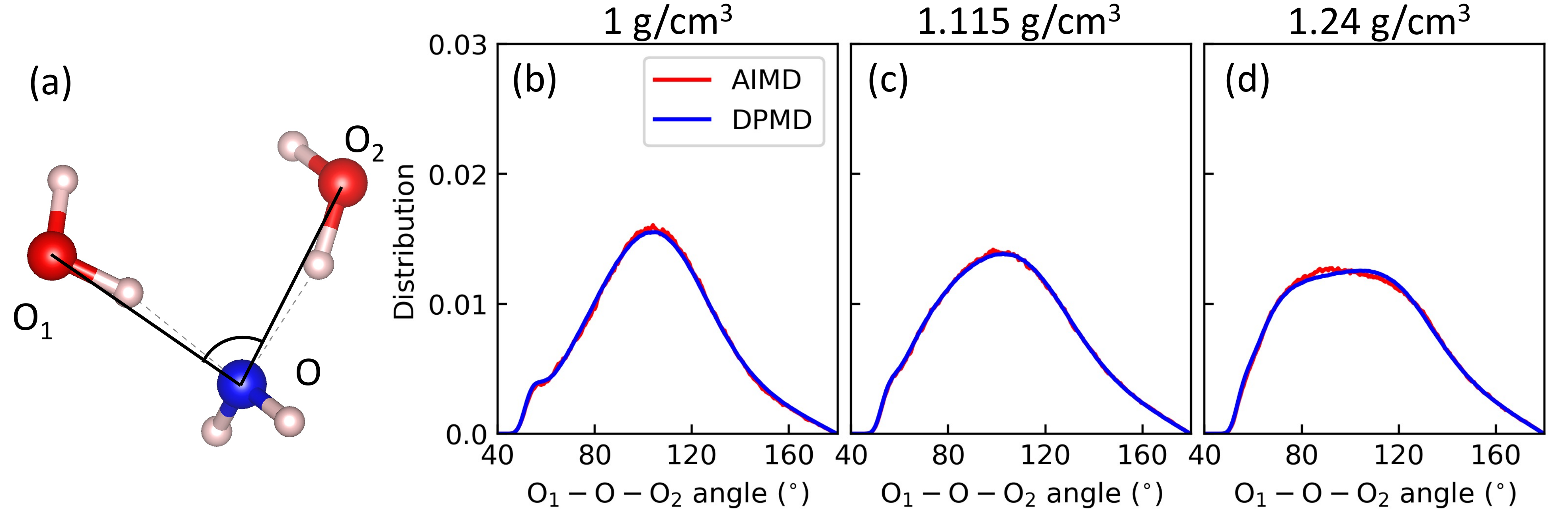}
  \caption{
  (a) Illustration of the O-O-O angle, where $\mathrm{O_1}$-$\mathrm{O}$ and $\mathrm{O_2}$-$\mathrm{O}$ distances are less than a certain $\mathrm{r_{cut}}$.
  (b-d) Comparison of the angular distribution function of O-O-O angle between AIMD (red) and DPMD (blue) at 1, 1.115, and 1.24 \cc{}.
  }\label{adf_compare}
 \end{figure}

In order to verify the accuracy of the trained DPMD models for liquid water, we compare the radial distribution functions (RDFs) and angle distribution functions (ADFs) as obtained from the AIMD and DPMD trajectories.
Fig.~\ref{pdf_compare} shows the RDFs including $\mathrm{g_{OO}(r)}$, $\mathrm{g_{OH}(r)}$ and $\mathrm{g_{HH}(r)}$ at density of 1.0, 1.115 and 1.24 \cc{}. The results are obtained from AIMD and DPMD trajectories.
We observe that the RDFs evaluated from the two trajectories match well.
Fig.~\ref{adf_compare} illustrates the O-O-O ADF of liquid water at the three densities; 
here the O-O-O angle refers to the angle formed by an O atom in a water molecule in the center and two of its neighbors.
In order to identify the first peak in the RDF, we set the neighboring cutoff radius to 3.215, 3.115, and 3.035 $\mathrm{\AA}$ for the 1, 1.115, and 1.24 \cc{} systems, respectively.
In fact, we find the resulting O-O-O distribution is relatively insensitive to the choice of the cutoff.
The above tests suggest that the AIMD and DPMD trajectories yield quantitatively comparable results,
suggesting that the structural features from AIMD are well preserved through the training of the DPMD models.

\section{Radial Distribution Functions and Static Structure Factors}
The manuscript has compared the computed RDFs and SSFs with the experiments in Fig.~1(e-g).~\cite{13JCP-Skinner, 16JCP-Skinner, 12JML-Yamaguchi}
The RDFs and SSFs data are obtained at three water densities (1, 1.115, and 1.24 \cc{}) from AIMD and DPMD trajectories.
At the densities of 1 and 1.115 \cc{}, the SSFs of $\mathrm{S_{OO}}$ are shown. However,
at the density of 1.24 \cc{}, the total SSF is shown and compared since $\mathrm{S_{OO}}$ is not provided in Ref.~\citenum{12JML-Yamaguchi}.

For all three densities, both RDF and SSF show reasonable agreement with the experiment.
We also notice that the first peak of $\mathrm{g_{OO}}(r)$ is slightly higher than the experimental results,
indicating a slightly over-binding hydrogen bond (HB) described by the SCAN functional.
To generate the full SSF of water,
the form factors of the O and H atom are set to be fractional to the electron number in accordance with Ref.~\citenum{12JML-Yamaguchi}, which is $\mathrm{f_O: f_H=8:1}$.

\RX{In addition, in order to compare the radial distribution functions at different densities, we replotted the RDF $\mathrm{g_{OO}}(r)$ and the decomposed RDF normalized by ambient density.
In practice, we multiply the RDFs with a normalization factor of the bulk density in Fig.~\ref{pdf_resize}.
After removing the bulk density from the denominator, the normalized RDF directly displays the absolute density at a certain distance from the central water molecule.
As a result, the three peaks in Fig. S4(b) match well, suggesting that the first-shell radial distribution functions contributed by HBs are almost unchanged for the three studied densities.
The conclusion is also consistent with the similar mean HB number of around 3.6 at the three densities (3.59 for 1 \cc{}, 3.58 for 1.115 \cc{}, 3.62 for 1.24 \cc{}).
Thus, the increase of the first peak in Fig.~\ref{pdf_resize}(a) is due to the invasion of the non-H-bonded water molecules into the inner shell, as shown by the onset of non-H-bonded radial distribution in Fig.~\ref{pdf_resize}(c).
}


\section{Hydrogen Bond Criteria}

\begin{figure}[htp]
  \includegraphics[width=14cm]{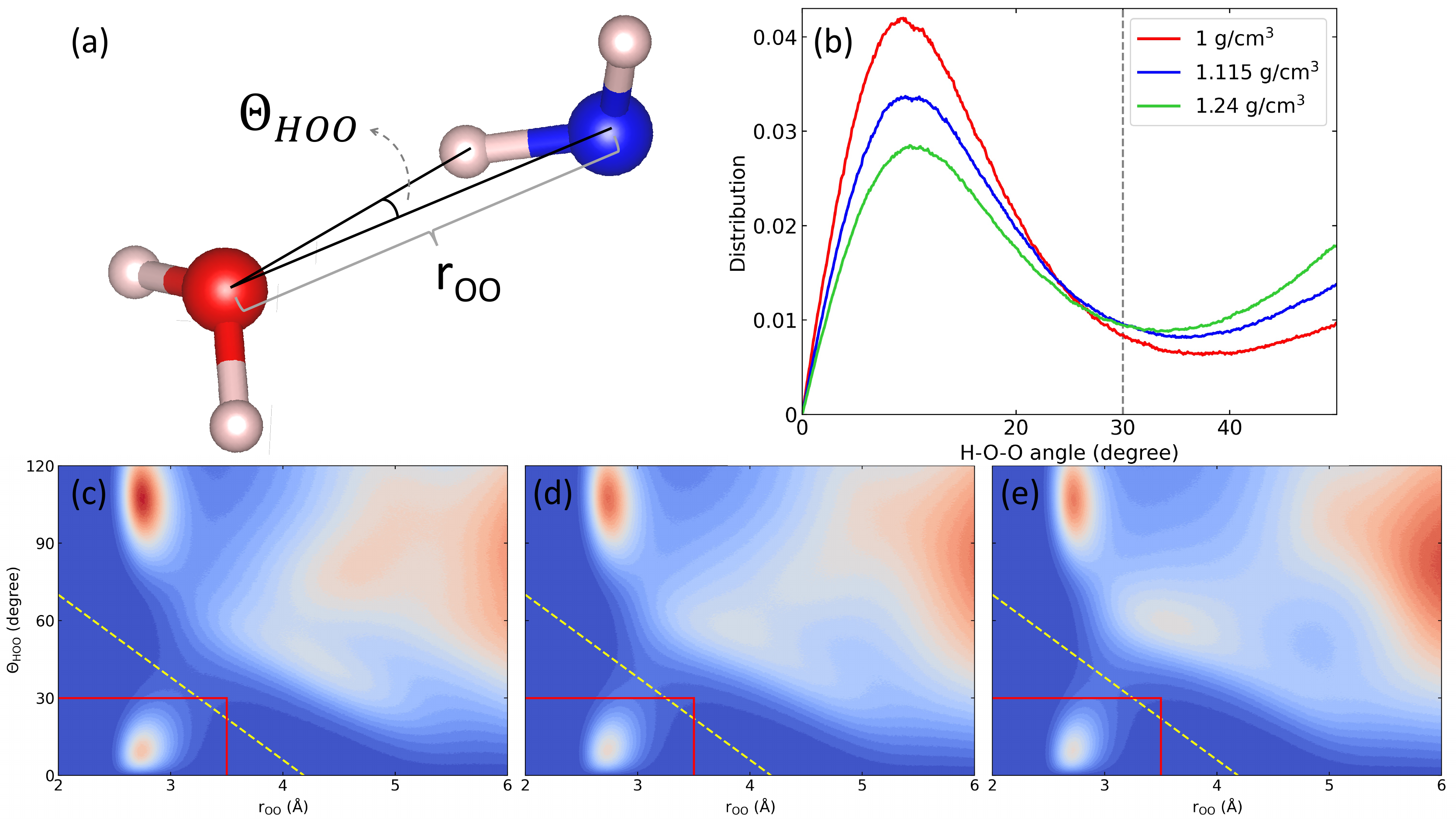}
  \caption{
 (a) A sketch illustrating the two values, $\mathrm{r_{OO}}$ and $\mathrm{\theta_{HOO}}$ involved in the definition of HB.
 (b) Distribution of $\mathrm{\theta_{HOO}}$ at densities of 1, 1.115, 1.24 $\mathrm{g/cm^3}$.
 (c-e) Joint distribution of $\mathrm{r_{OO}}$ and $\mathrm{\theta_{HOO}}$ at densities of 1, 1.115, 1.24 $\mathrm{g/cm^3}$.
 The adopted HB definition in the paper and a looser definition are respectively marked with red and yellow lines.
  }\label{HB_def}
 \end{figure}

Different definitions of HB exist for both ambient water and water under pressure.
As elucidated in the manuscript and previous literature, ranging from ambient pressure to 1 GPa,
the influence of high pressure on hydrogen bonds remains limited. In this regard, we keep the classical HB classification standard to facilitate comparison with earlier literature.
This is particularly true considering that the discussion in the manuscript only involves structural changes rather than dynamic changes.
However, it is still necessary to examine the sensitivity of our conclusions against changes in the HB criterion.

\begin{figure}[htp]
  \includegraphics[width=14cm]{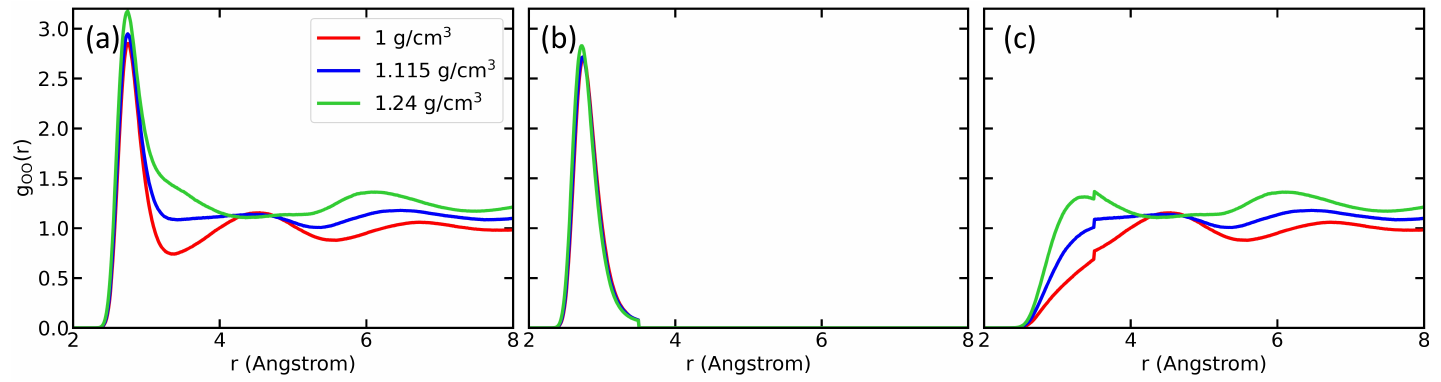}
  \caption{\RX{
 (a-c) Radial distribution function (RDF) $\mathrm{g_{OO}(r)}$, RDF contributed by hydrogen-bonded water molecule pairs, and RDF contributed by non-hydrogen-bonded water molecule pairs normalized by the density of ambient water. Water systems with a density of 1 \cc{}, 1.115 \cc{}, and 1.24 \cc{} are respectively colored in red, blue, and green.}
  }\label{pdf_resize}
 \end{figure}

Figs.~\ref{HB_def}(c-e) show the joint distribution of $\mathrm{\theta_{HOO}}$ angle and $\mathrm{r_{OO}}$ between two water molecules (Fig.~\ref{HB_def}(a)).
At all three densities,
the H-bonded water pairs clearly form a relatively independent single-peak distribution separate from non-H-bonded water at the lower left part,
indicating that the HB distribution is barely changed by pressure under 1 GPa.
The vertical line and horizontal line in red represent $\mathrm{r_{OO}=3.5~\AA}$ and $\mathrm{\theta_{HOO}=30^{\circ}}$, respectively. The two red lines encircle the classical H-bonded distribution area defined by Luzar et al.~\cite{96N-Luzar}
As shown by the joint distribution and $\mathrm{g_{OO}(r)}$ in Figs. 2(d-e) of the manuscript,
the radial cutoff of 3.5 $\mathrm{\AA}$ is a safe choice.
Nevertheless, the angular cutoff of 30$\mathrm{^{\circ}}$ on $\mathrm{\theta_{HOO}}$ could be a strict criterion and might oversee a few marginal HB cases, as shown in Fig.~\ref{HB_def}(b).

How to classify hydrogen bonds more accurately is another issue that requires different perspectives to justify. In this work, we choose to draw a line on the joint distribution as a loose upper bound of HB to include as many (marginal) HB cases as possible.
The line chosen here is $\mathrm{\theta_{HOO} (^{\circ})=-32 \times r_{OO} (\AA) + 134}$, marked as yellow broken line in Figs.~\ref{HB_def}(c-e).
In the following part, we repeat the analyses related to the HB standard in the manuscript with this looser HB standard to justify the robustness of the conclusions in the manuscript.
The looser HB standard used here would be referred to as the new HB standard in the following parts.

\begin{figure}[htp]
  \includegraphics[width=14cm]{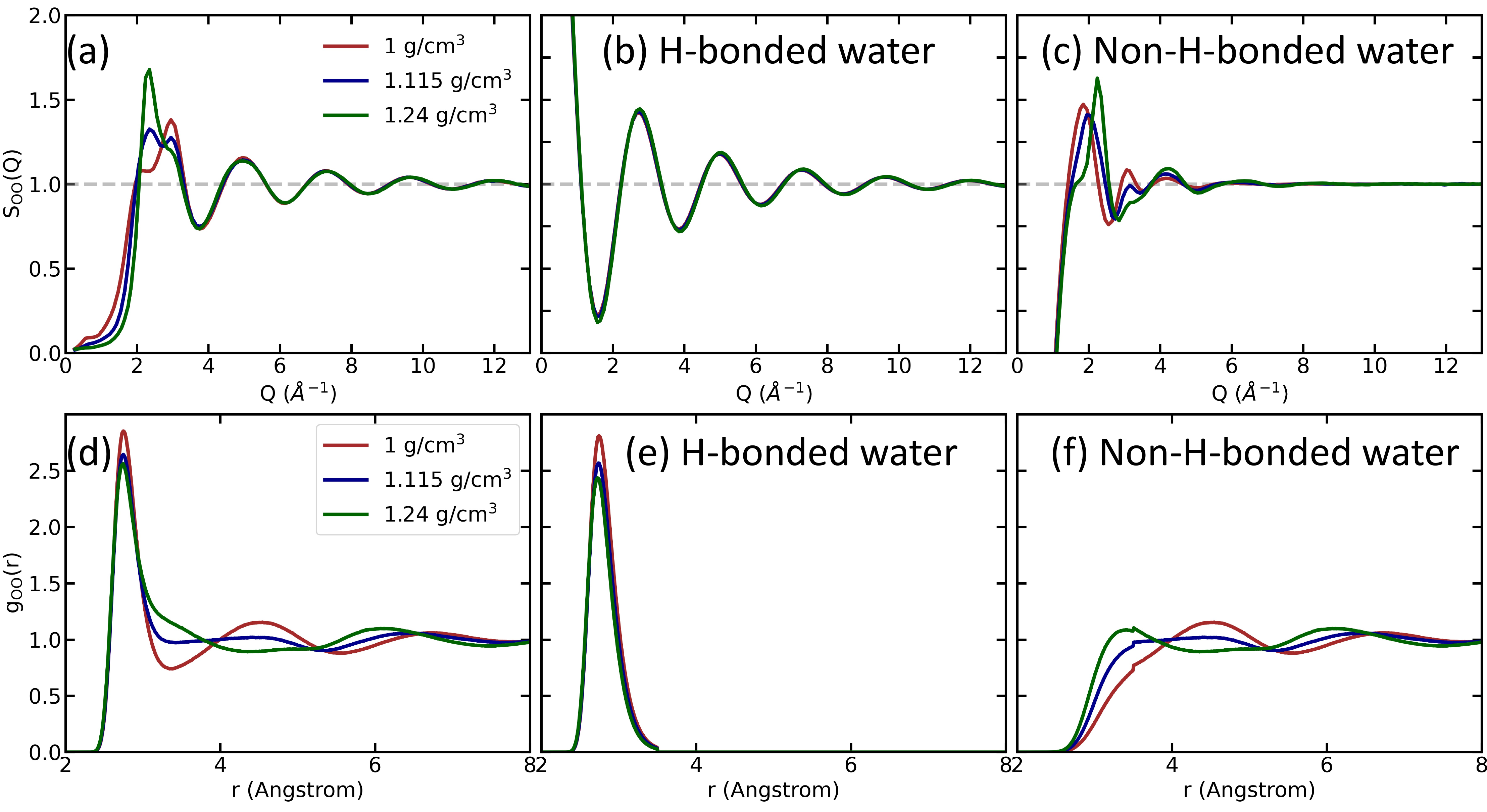}
  \caption{
  (a-c) Static structure factor (SSF) $\mathrm{S_{OO}(q)}$, SSF contributed by hydrogen-bonded water molecule pairs and SSF contributed by non-hydrogen-bonded water molecule pairs. (d-f) Radial distribution function (RDF) $\mathrm{g_{OO}(r)}$, RDF contributed by hydrogen-bonded water molecule pairs, and RDF contributed by non-hydrogen-bonded water molecule pairs. Water systems with a density of 1 \cc{}, 1.115 \cc{}, and 1.24 \cc{} are respectively colored in red, blue, and green.
  The new HB standard (yellow broken line in Fig.~\ref{HB_def}) is used herein.
  }\label{pdf_ssf}
 \end{figure}

\begin{figure}[htp]
  \includegraphics[width=10cm]{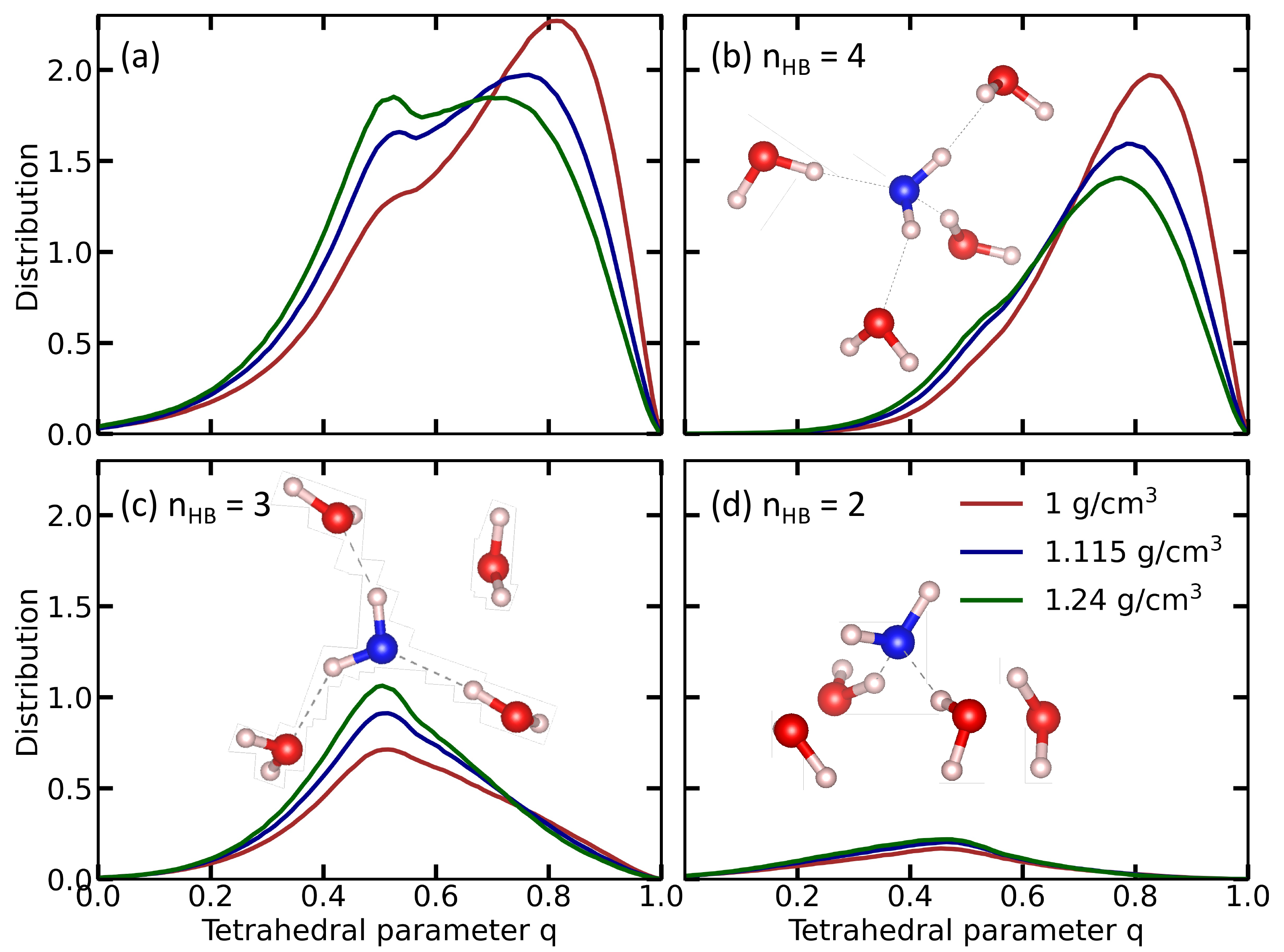}
  \caption{
 (a) Distribution of tetrahedral parameter q of water under three different pressures.
 (b-d) Distributions of $q$ contributed by water molecules whose 4, 3, and 2 neighbors out of the four closest neighbors are H-bonded, denoted by $\mathrm{n_{HB}=4, 3, 2}$, respectively. 
The HBs are classified using the new standard here (yellow broken line in Fig.~\ref{HB_def}).
  Inset in each subplot depicts a typical water with its four closest neighbors of the corresponding $\mathrm{n_{HB}}$.
  }\label{tetra_q}
 \end{figure}

\begin{figure}[htp]
  \includegraphics[width=8.5cm]{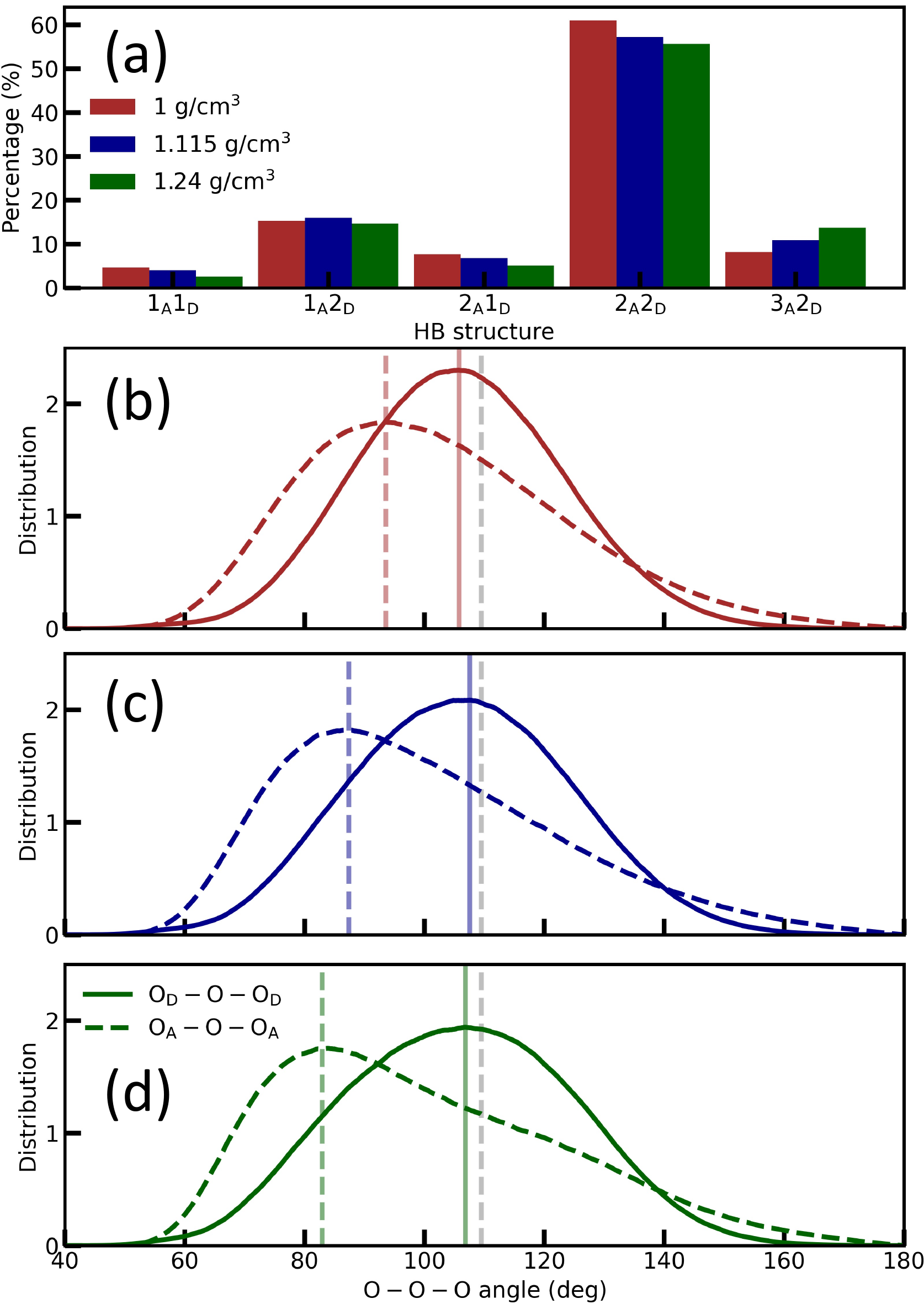}
  \caption{
 (a) Percentages of five major HB structures observed in liquid water with densities being 1, 1.115, and 1.24 \cc{} under the new HB standard.
  (b-d) Distributions of the $\mathrm{O_A}$-$\mathrm{O}$-$\mathrm{O_A}$ angle (broken line) and $\mathrm{O_D}$-$\mathrm{O}$-$\mathrm{O_D}$ angle (solid line) of the $\mathrm{2_A2_D}$ structure at the three different densities under the new HB standard.
  Vertical broken and solid lines in corresponding colors mark the peak positions of the $\mathrm{O_A}$-$\mathrm{O}$-$\mathrm{O_A}$ and $\mathrm{O_D}$-$\mathrm{O}$-$\mathrm{O_D}$ angle distribution, while the grey broken line mark 109$^{\circ}$28$^{\prime}$.}\label{HB_stru_stat}
 \end{figure}

We replotted Figs.~2-4 in the manuscript with the new HB standard,
which are shown as Figs.~\ref{pdf_ssf}-\ref{HB_stru_stat}.
First, for the one-to-one decomposition of RDF and SSF, the new HB standard yields decomposed results (Fig.~\ref{pdf_ssf}) that are basically the same as Fig.~2 in the manuscript,
so the SSF changes with respect to increasing pressure are undoubtedly related to changes in the structure of non-H-bonded water molecules.

Second, the tetrahedral order parameter $q$ distribution and its decomposition with respect to $\mathrm{n_{HB}}$ are shown in Fig.~\ref{tetra_q}.
With the new HB definition,
the change in the order parameter distribution could still be attributed to the decrease in the $\mathrm{n_{HB}=4}$ proportion and increase in the $\mathrm{n_{HB}=2, 3}$ proportion.
However, the distribution of $q$ of $\mathrm{n_{HB}=3, 4}$ also slightly changes as the HB standard changes.
The $q$ distribution of $\mathrm{n_{HB}=4}$ between 0.4 and 0.6 slightly increases, forming a bump therein at the densities of 1.115 and 1 $\mathrm{g/cm^3}$.
Meanwhile, distribution of $\mathrm{n_{HB}=3}$ between 0.4 and 0.6 decreases,
resulting in a single-peak shape in closed resemblance to $\mathrm{n_{HB}=2}$ distribution.
This indicates that the loose HB standard classifies a number of $\mathrm{n_{HB}=3}$ cases under the classical standard into the $\mathrm{n_{HB}=4}$ class.
The order parameter $q$ of newly added $\mathrm{n_{HB}=4}$ cases are mostly between 0.4 and 0.6,
indicating a poor tetrahedrality in such coordination.
The change is not surprising since the new standard classifies more marginal HB into HB, 
which are mostly on the boundary between H-bonded and non-H-bonded water pairs.
These marginal HB are very likely to belong to not-so-tetrahedral HB structures.

Third, Fig.~\ref{HB_stru_stat} shows the content in Fig.~4 of the manuscript with the new HB standard.
For the HB structure distribution,
the percentage of $\mathrm{2_A2_D}$ drops monotonically as percentage of $\mathrm{3_A2_D}$ increases as pressure increases, 
same as the results from classical standards.
But the percentage of $\mathrm{1_A2_D}$ does not vary much as pressure increases under the new standard.
The new HB standard increases the percentage of $\mathrm{2_A2_D}$ from 54.2\%, 49.4\%, 48.7\% under the classical standard to 61.0\%, 57.2\%, and 55.7\%.
This dramatic change, 
together with the change in the shape of $\mathrm{n_{HB}=4}$ q distribution, 
indicates that a number of marginal HBs belong to $\mathrm{2_A2_D}$ structures with poor tetrahedrality.
Furthermore, the percentage of HB structure with 3 donated HB (not shown in Fig.~\ref{HB_stru_stat}) also increases from 0.5\%, 0.8\%, and 1\% under the classical standard to 1.6\%, 3.1\%, and 5.6\%.
Since most 3-donated HB cases are bifurcated HB from `HB jump'\cite{06S-Laage},
the increase shows that HB jump is another important source of marginal HB,
and the frequency of HB jump grows as pressure increases.
The increase in frequency might be a sign of the growing instability of donated HB as pressure increases.
As for the $\mathrm{O}$-$\mathrm{O}$-$\mathrm{O}$ angle distribution,
the new HB standard yields similar results to the classical standard:
$\mathrm{O_A}$-$\mathrm{O}$-$\mathrm{O_A}$ angle decreases as pressure increases while $\mathrm{O_D}$-$\mathrm{O}$-$\mathrm{O_D}$ angle distribution basically remains unchanged by the pressure.

In conclusion,
switching the HB standard to a looser one generally does not change the conclusions related to HB classification.
A looser HB bound also reveals a few new understandings concerning HB structure.
For example, the marginal HBs mostly come from highly distorted $\mathrm{2_A2_D}$ structure or the process of HB jump, i.e. $\mathrm{3_D}$ structures.
The increase in the $\mathrm{3_D}$ structures at a higher pressure indicates a growing frequency of HB jump as pressure increases,
which requires further study in the future.

  \begin{figure}[htp]
  \includegraphics[width=14cm]{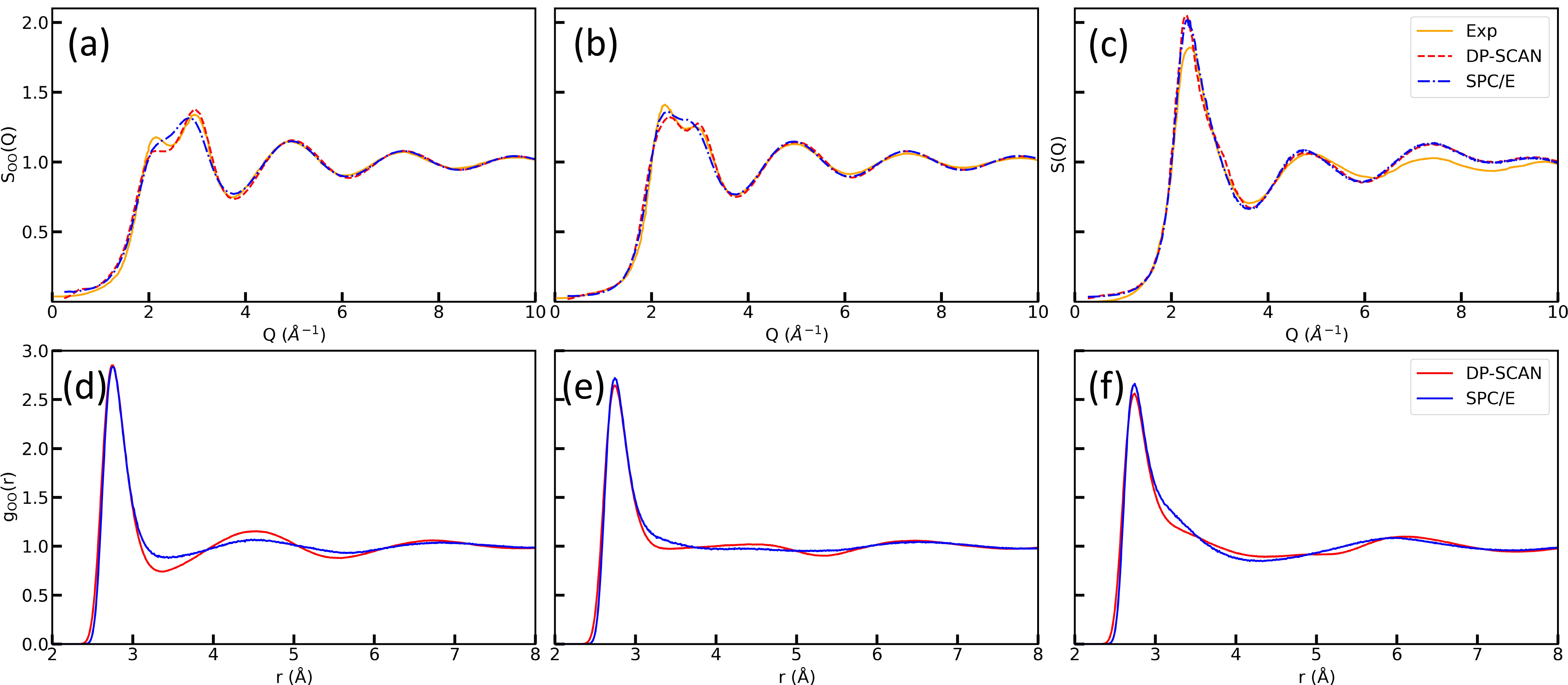}
  \caption{\RX{
  (a-c) SSFs from experiment, DPMD and classical MD with SPC/E empirical potential at density of 1 \cc{}, 1.115 \cc{} and 1.24 \cc{}. For 1 \cc{} and 1.115 \cc{}, $\mathrm{S_{OO}(Q)}$ is shown. For 1.24 \cc{}, a total SSF S(Q) is shown to facilitate comparison with experiment. Results from experiment, DPMD and classical MD are shown in orange, red and blue curves respectively.
 (d-f) RDF $\mathrm{g_{OO}(r)}$ from DPMD and classical MD with SPC/E empirical potential at density of 1 \cc{}, 1.115 \cc{} and 1.24 \cc{}. Colors of the RDF curves are consistent with the SSF.}
 }\label{spce}
 \end{figure}

\section{Definition of Tetrahedral Order Parameter}
The parameter is defined in consistency with Ref.~\citenum{01N-Errington} as $q=1-\frac{3}{8}\sum_{j=1}^{3}\sum_{k=j+1}^{4}(\cos\psi_{jk}+\frac{1}{3})^2$, 
where $\psi_{jk}$ denotes the angle formed by the O atom in a water molecule and two O atoms from its four closest neighbors, 
in the same way, as shown in Fig.~\ref{adf_compare}(a).
For standard tetrahedron, $\psi_{jk}=-1/3$ for all $j\neq k$, thereby rendering $q=1$.
As is stated in Ref.~\citenum{01N-Errington}, for purely random cases like an ideal gas, the ensemble average of $q$ is an integration over $\psi_{jk}$ from $0$ to $\pi$, 
rendering $\langle q \rangle = 0$.
However, in our simulated system, we find that a structure with $q=0.5$ could already be far from a standard tetrahedral structure,
as is shown in the inset of Fig. 3(b).

\section{Comparison with Empirical Potential}

\RX{We compare the results from DPMD and classical MD using empirical potential.
We performed classical MD with SPC/E model~\cite{87JPC-Berendson} in LAMMPS software.
All other MD parameters were set in consistency with the DPMD.
The resulting SSFs and RDFs in comparison with DPMD and experiment are shown in Fig.~\ref{spce}.
}

\RX{In general, the SPC/E model qualitatively captures the major structural change induced by pressure,  i.e., the reduction in the second peak of RDF $\mathrm{g_{OO}(r)}$ and the rise in the right shoulder of the first peak.
The result is also in consistency with previous studies using empirical models.~\cite{97JCP-Bagchi, 03E-Saitta}
However, we note that SPC/E fails to reproduce the SSF of liquid water quantitatively at density of 1 \cc{} and 1.115 \cc{}.
The first two peaks at the two densities degrade into a biased peak. 
As shown in the Fig.~\ref{spce}(d-e), the mismatch mainly comes from a poor description of the water in the interstitial region and beyond.
As for the density of 1.24 \cc{}, the SSFs from SPC/E model and DPMD are similar,
suggesting that the structural description of classical model tends to be more accurate as pressure increases, pushing the structure of water towards simple liquid.
}

  \begin{figure}[htp]
  \includegraphics[width=12cm]{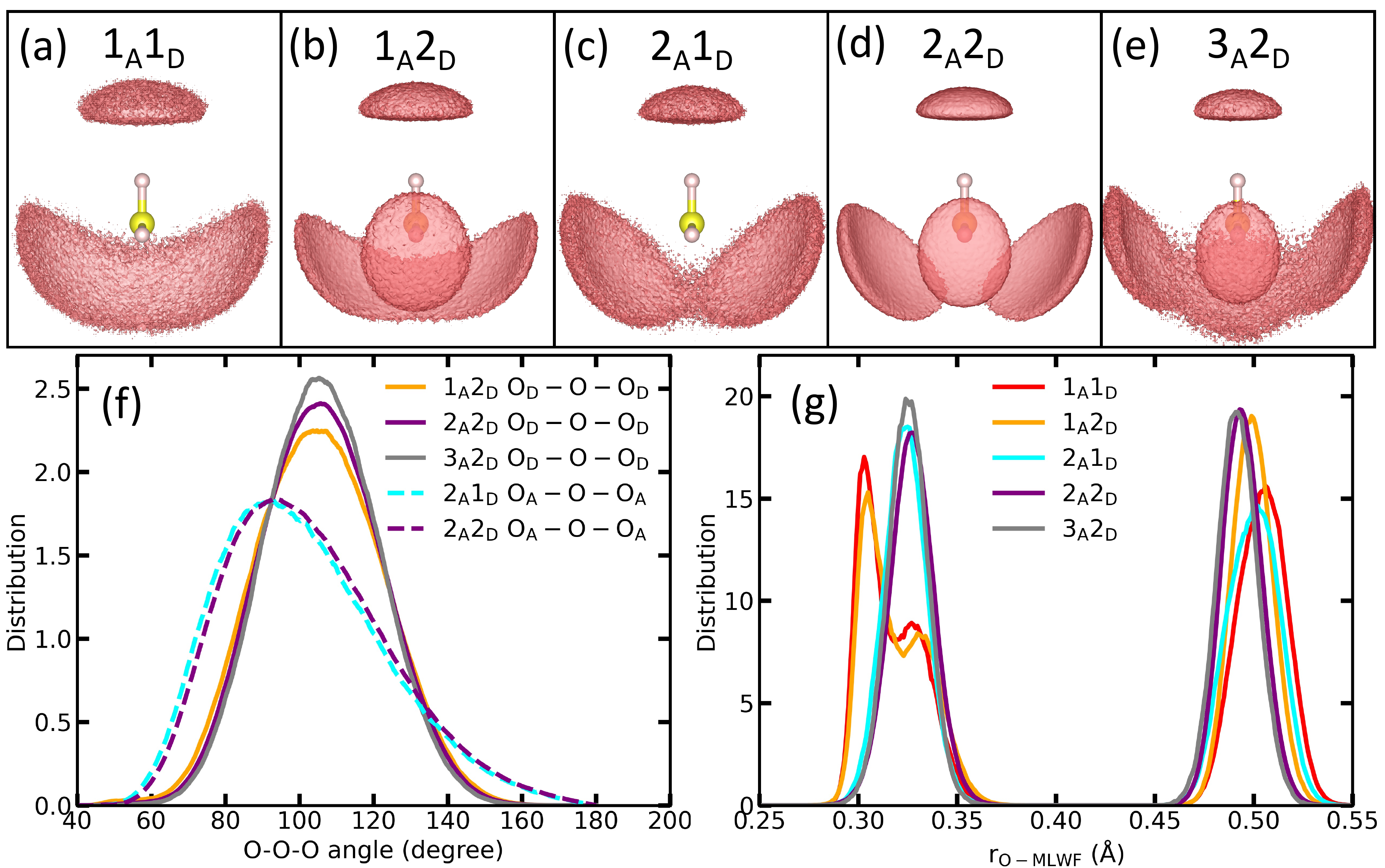}
  \caption{
 (a-e) Spatial distribution function (SDF) of O atom in the first-shell water of the five major HB structures under ambient pressure.
 (f) Distribution of $\mathrm{O_D}$-$\mathrm{O}$-$\mathrm{O_D}$ angle of $\mathrm{1_A2_D}$, $\mathrm{2_A2_D}$ and $\mathrm{3_A2_D}$ HB structures, and $\mathrm{O_A}$-$\mathrm{O}$-$\mathrm{O_A}$ angle of $\mathrm{2_A1_D}$ and $\mathrm{2_A2_D}$ HB structures. 
 (g) Distribution of distance between the Wannier center and the O atom of a water molecule.}\label{wannier}
 \end{figure}

\section{Electronic Structure}
In the last section, we present our new findings about the relation between the HB structure and the HB strength.
The result is almost unaffected by the density range studied,
so only results from ambient water are provided here.
As shown in Figs.~\ref{wannier}(a-e), 
not only the shape but also the spread range of spatial distribution of HB depend on the HB structure.
Comparing the donating ends in Figs.~\ref{wannier}(b), (d) and (e), 
the donating HBs of a water molecule exhibit more localized distribution as the water accepts more HBs.
Similarly, comparing the accepting ends in Fig.~\ref{wannier}(a) and (b), (c) and (d),
the more HBs a water molecule donates,
the more localized its accepting HB distribution is observed.
The phenomenon is also revealed by the O-O-O angle distribution of different HB structures:
the spread of $\mathrm{O_D}$-$\mathrm{O}$-$\mathrm{O_D}$ angle distribution grows larger monotonically from $\mathrm{3_A2_D}$, $\mathrm{2_A2_D}$ to $\mathrm{1_A2_D}$.
Meanwhile, the $\mathrm{O_A}$-$\mathrm{O}$-$\mathrm{O_A}$ angle distribution of $\mathrm{2_A1_D}$ moves slightly leftward compared to that of $\mathrm{2_A2_D}$.
The phenomenon reminds us of the dependence of accepted (donated) HB number distribution on donated (accepted) HB number,~\cite{08MP-Markovitch}
which suggests that the water tends to accept (donate) more HBs as it donates (accepts) more HBs.
All of the trends indicate that the strength of HB is relevant to the HB number on the other side of the water molecule,
i.e., the strength of the accepted (donated) HB of a water molecule grows as the water donates (accepts) more HB. 
%

In view of this, we examine the dependence of electronic structure on the HB structure.
Since the electronic structure of the water molecule can be effectively represented by maximally localized Wannier function (MLWF)~\cite{97B-Marzari, 12RMP-Marzari}, 
we compare the electronic structure of the major HB structures by examining their MLWF distributions.
Fig.~\ref{wannier}(g) displays the distribution of the distance of the center of MLWF from the oxygen atom (short as $\mathrm{r_{O-MLWF}}$ in the following) in each HB structure at 1 \cc{},
where the left peak represents the lone pair, and the right peak represents the bonding pair.
We also check the MLWF distributions at different pressures,
and find out that the influence of pressure on Wannier function distribution is quite small.
Notably, we find that the $\mathrm{r_{O-MLWF}}$ distributions are rather different in different HB structures (Figs.~\ref{wannier}(a-e)).
For instance, the lone pairs of $\mathrm{1_A}$ structures form two peaks, the right one of which locates at the same position as $\mathrm{2_A}$ structures.
The separation of peaks of $\mathrm{1_A}$ structure suggests that the acceptance of one HB could directly pull the lone pair away from the O atom while leaving the other lone pair relatively unchanged.
On the donating end, the separation of peaks is not that strong; however,
we observe a similar situation:
donating an HB could push the corresponding bonding pair towards the O atom while leaving the other pair unmoved.
Furthermore, 
we focus on the dependence of the lone pair (bonding pair) on the number of donated (accepted) HB.
We find that
the distance between the lone (bonding) pairs and the O (H) atom becomes larger as the water molecule donates (accepts) more HBs.
A larger distance from the lone pair to the O atom stands for a more negative environment at the O end,
thereby enabling stronger HB to form at the accepting end.
On the contrary,
a larger distance from the bonding pair to the H atom stands for a more positive environment at the H end,
enabling stronger HB to form at the donating end.
The above analysis explains the HB number, the directionality, and the strength of one end of a water molecule depending on the HB number on the other end.

%
%
%
%

\bibliography{SM_ref}